\documentclass[conference]{IEEEtran}

\makeatletter
\def\ps@headings{%
\def\@oddhead{\mbox{}\scriptsize\rightmark \hfil \thepage}%
\def\@evenhead{\scriptsize\thepage \hfil \leftmark\mbox{}}%
\def\@oddfoot{}%
\def\@evenfoot{}}
\makeatother
\usepackage{graphics, epstopdf, graphicx, epsfig, psfrag, subfigure}
\usepackage{dcolumn, multirow}%
\usepackage{verbatim}
\usepackage{amsmath,  amssymb, latexsym,amsfonts,epsfig, graphicx, ,bm}
\usepackage[nospace]{cite}
 \usepackage{algorithmic, algorithm}
\usepackage{verbatim}
\usepackage{xspace}
\usepackage[table]{xcolor}
\usepackage{color, colortbl}
\usepackage{tikz}
\usetikzlibrary{shapes,arrows}

\usepackage{graphicx}%
\usepackage{dcolumn}%
\usepackage{bm}%
\usepackage{epsfig}
\usepackage{psfrag}
\usepackage{subfigure}
\usepackage{color}
\usepackage{amsmath, amssymb, latexsym,amsfonts,epsfig, graphicx, verbatim}
\usepackage[nospace]{cite}
\usepackage{verbatim}
\usepackage{xspace}

\newcommand{\eqn}[1]{Eq.(\ref{#1})}
\newcommand{\Sec}[1]{Sec.~\ref{#1}}
\newcommand{\Fig}[1]{Fig.~\ref{#1}}
\newcommand{\Tab}[1]{Table~\ref{#1}}

\newcommand{\smaller}{\!<\!}

\newcommand{\w}{\textrm{w}}

\newcommand{\RW}{{\scriptscriptstyle\textrm{RW}}}

\newcommand{\sss}[1]{{\scriptscriptstyle\textrm{#1}}}

\newcommand{\FOR}{\textbf{for }}
\newcommand{\FORALL}{\textbf{forall }}
\newcommand{\DO}{\textbf{ do }}

\newcommand{\IF}{\textbf{if }}
\newcommand{\AND}{\textbf{and }}

\newcommand{\THEN}{\textbf{ then }}

\newcommand{\REQUIRE}{\textbf{require }}

\newcommand{\SP}{${}$\hspace{0.5cm}}

\newcommand{\est}[1]{\widehat{#1}}

\newcommand{\ie}{{\em i.e., }}
\newcommand{\eg}{{\em e.g., }}

\newcommand{\dK}{$dK$\xspace} 

\newcommand{\Count}{\textrm{count}}
\newcommand{\JDD}{\textrm{\small{JDD}}\xspace}

\newcommand{\target}[1]{#1^\odot}
\newcommand{\CCK}{$\bar{c}(k)$\xspace}

\begin{document}

\title{2.5K-Graphs: from Sampling to Generation}

\begin{comment}
\author{Minas Gjoka$\dag$, Maciej Kurant$\ddag$, Athina Markopoulou$\dag$\\
$\dag$ CalIT2,  University of California, Irvine\\ %
$\ddag$ ETH, Zurich %
}
\end{comment}

\author{Minas Gjoka\\
CalIT2,  University of California, Irvine\\ 
{\tt mgjoka@uci.edu}\\
\and
Maciej Kurant\\
ETH Zurich\\
{\tt maciej.kurant@gmail.com}\\
\and
Athina Markopoulou\\
EECS and CalIT2, UC Irvine\\
{\tt athina@uci.edu}\\
}

\maketitle

\begin{abstract}
Understanding network structure and having access to realistic graphs plays a central role in  computer and social  networks research.  
 In this paper, we propose a complete, and practical methodology for generating graphs that resemble a real graph of interest.  The metrics of the original topology we target to match are the joint degree distribution (JDD) and the degree-dependent average clustering coefficient  ($\bar{c}(k)$). We start by developing efficient estimators for these two metrics based on a node sample collected via either independence sampling or random walks. Then, we process the output of the estimators to ensure that the target properties are realizable. Finally, we propose an efficient algorithm for generating topologies that have the exact  target JDD and a $\bar{c}(k)$ close to the target. Extensive simulations using real-life graphs show that the graphs generated by our methodology are similar to the original graph with respect to, not only the two target metrics,  but also a wide range of other topological metrics; furthermore, our generator is order of magnitudes faster  than state-of-the-art  techniques.

\begin{comment}
In this paper we propose a practical algorithm to generate topologies with an exact joint degree distributions and degree dependent clustering coefficient distribution. We show in simulations that it can be orders of magnitude faster that previous approaches.  \textcolor{red}{Our key insight is that destroying triangles is easier than adding them}. We use simulations in real-life graphs to show that the generated graphs are similar to the original graph with respect to a wide range of topological metrics. Last, we release  a tool that implements our algorithm.
\end{comment}

\end{abstract}

\section{Introduction}\label{sec:intro}

Understanding network structure and having access to realistic graphs plays a central role in  computer and social  networks research.  
To this end, researchers use various approaches: they collect real measurements, often involving sampling; use the datasets for a  particular study and/or make them publicly available; and develop analytical models that can generate topologies with key properties resembling those of the network of interest. Each of these approaches is a challenging research question on its own,  involving complex tradeoffs between accurate representation of the original graph and practical constraints in terms of measurement overhead, algorithm complexity, privacy.

For example, the popularity of online social networks  (OSNs)
 has given rise to a number of measurement studies to improve the understanding of their characteristics. Being able to characterize and simulate the  topology of the social graph is  important for evaluating the effectiveness of a growing number of social network applications that attempt to leverage the social graph. The commonly used approach is to measure these networks and make the dataset available and properly anonymized. Given the size of most of these networks, they are not typically measured in their entirety, instead \emph{sampling} is used to estimate properties of interest. 
 Another approach is to develop models that allow to generate graphs that meet certain properties of interest, such as node degree distribution, clustering coefficient, community structure, diameter, etc. 

So far, network sampling and topology generation have been looked at separately.  We believe there is a need for a complete  methodology that starts by sampling a real (yet not fully known) graph, estimates properties of interest, and generates synthetic graphs that resemble the original in a number of important properties; the methodology should also meet practical constraints such as  sampling budget and computational complexity.

\begin{comment}
The usage of OSNs is quite popular nowadays, with multiple types of social networks available for various human activities. To mention a few examples, Facebook is used to connect with friends, Twitter for microblogging, Linkedin to connect with colleagues and classmates, Instagram to broadcast pictures etc. The number of registered users in the most popular OSNS has risen to the order of hundreds of millions.

Recent measurement studies of OSNs have provided a first step towards their understanding. So far, a number of works \cite{Gjoka2010,Ribeiro2010,Gjoka2011_multigraph_JSAC,Kurant2011_SWRW} have succesfully used principled methods to gather representative node samples and measure local properties of the social graph. However, little has been done on the measurement of non-local properties that are related with the network structure. Being able to measure and simulate such network structure is quite important in evaluating the effectiveness of a growing number of social network applications that attempt to leverage the social graph. The commonly used approach for such evaluations is to collect or use a Breadth-First-Search sample of the social graph since it provides a full view between the sampled  users. However such samples are inherently biased even in simple graph properties such as degree, let alone network structure. We believe there is a need for a methodology that, starting from node samples, can  produce synthetic graphs that resemble the original graph, in a number of important properties.

\end{comment}

There is, of course, a plethora of metrics one could be interested in when analyzing and generating graphs. Ideally, one would like to generate synthetic graphs that resemble the original in as many  topological properties as possible.   %\footnote{%
In practice, there are two main limitations.  First, the topological properties of interest should be estimated based on a  sample of the real graph; the more involved the properties, the larger sample is needed.
Second, constructing a graph with given properties becomes more difficult for more restrictive properties.

Given these practical limitations in estimation and graph generation, we focus on the following two metrics:  joint degree distribution, \JDD,  and degree-dependent average clustering coefficient, \CCK. We develop a complete, practical methodology for generating graphs that resemble a real graph of interest in terms of these two metrics. 

We follow the systematic framework and terminology of dK-series \cite{Mahadevan2006a}, which characterizes the properties of a graph using series of probability distributions specifying all degree correlations within d-sized subgraphs of a given graph G.  Increasing values of d capture progressively more properties of G at the cost of more complex representation of the probability distribution.
We refer to the graphs generated by our approach as {\em 2.5K-graphs}   because they meet more specified properties than  2K-graphs (\ie graphs with a target joint degree distribution) but less than 3K-graphs (graphs with specified distributions of all subgraphs of three nodes). 
\begin{comment}
 a family of graphs with a given $2K$ and \CCK. In other words, we propose to combine the $2K$ with some clustering information.  Because the full clustering information would be captured by $3K$, our proposal can be located somewhere between $2K$ and $3K$. For this reason, we refer to it as $2.5K$. 
%
We show that $2.5K$ strikes a good tradeoff between the level of details and applicability. 
\end{comment}
We show that  $2.5K$ provides a sweet spot between accurate representation and practical constraints. 
The key insight is that some information about clustering is necessary  for a realistic representation of real-life graphs, especially OSNs, while \CCK is still practical  to estimate and generate. More specifically, we make the following contributions  in estimation and generation  of 2.5K. 
\tikzstyle{block} = [rectangle, draw, fill=blue!20, text width=5em, text centered, rounded corners, minimum height=4em]

\tikzstyle{dot} = [rectangle, draw]
\tikzstyle{line} = [draw, -latex']

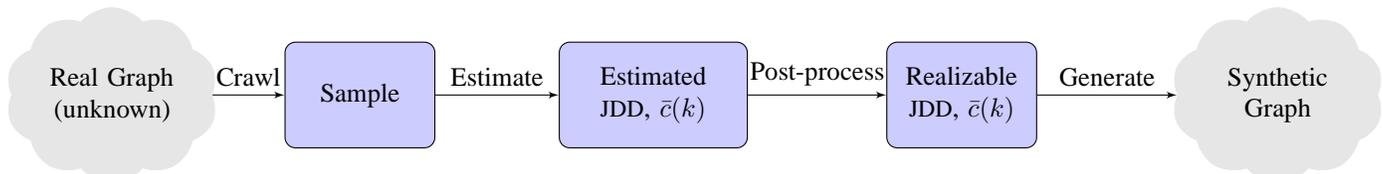
\begin{figure*}    
\begin{tikzpicture}[node distance = 2cm, auto]
    \node[cloud, fill=gray!20, cloud puffs=10, cloud puff arc= 100,
        text width=1.65cm, minimum height=1cm, aspect=1.3, align=center] (topology) {Real Graph \\ (unknown)};
    
    \node [block, right of=topology, node distance=3.3cm, align=center] (sample) {Sample};
    \node [block, node distance=3.9cm, right of=sample, minimum width=2.5cm, text width=2cm, align=center] (properties) {Estimated  \\ \JDD, $\bar{c}(k)$};
    \node [block, node distance=4.1cm,right of=properties, align=center] (realizable) {Realizable \\ \JDD, $\bar{c}(k)$};
    \node[cloud, node distance=4.2cm,right of=realizable, fill=gray!20, cloud puffs=10, cloud puff arc= 100, text width=1.65cm, minimum height=1cm, aspect=1.3, align=center] (synthetic) {Synthetic \\ Graph };

    \path [line] (topology) -- node {Crawl} (sample);
    \path [line] (sample) -- node {Estimate} (properties);
    \path [line] (properties) -- node {Post-process} (realizable);\\
    \path [line] (realizable) -- node  {Generate} (synthetic);
\end{tikzpicture}
\vspace{-12pt}
\caption{\label{fig:overview}Overview of our approach: Sampling, Estimation, Post-processing, Graph Generation. We start by sampling a large real (not fully known) graph (\eg Facebook) and we end by constructing synthetic graphs that are similar to the original, in terms of the 2.5K metrics (JDD and clustering). }
\vspace{-12pt}
\end{figure*}

{\em Estimation:} We derive efficient estimators of the metrics of interest, namely \JDD  and \CCK, based on  a node sample collected either via an independence sampler or via a random walk; the latter is the  common practice in sampling OSNs.  
Our design utilizes edges induced between sampled nodes, and appropriately corrects for biases introduced by random walk resulting from (i)~non-uniform sampling weights, and (ii)~strong dependencies between successive samples.
We demonstrate the efficiency of the estimators via simulation. In addition, we post-process these metrics to ensure that they are realizable \ie that there exist graphs with those properties. %

{\em Generation:}  We propose a practical algorithm for generating $2.5K$-graphs that follow the target \JDD exactly and \CCK very closely. Our algorithm starts by generating a graph with exactly \JDD and more triangles than needed; it then performs double edge swaps trying to meet the desired \CCK. The key intuition and novelty compared to prior approaches, is that destroying triangles is much easier than creating new ones. Extensive simulations of real graphs show several strengths of  our approach. First, the $2.5K$ synthetic graph~$G'$ is similar to the original~$G$ not only with respect to the targeted metrics, 
but also  to a wide range of other topological properties. %
Second, our generation algorithm is orders of magnitude faster than prior approaches; in fact, the latter may not converge in practice.
%
%

%
%

	%

%
%
%

%

%
%
%
%
%
%
%
%
%
%
%
%
%
%
%
%
%
%

\begin{comment}
\textcolor{red}{OSNs have high clustering. stress that achieving it is important}
%
$dK$-series specify degree correlations across all $d$-sized subgraphs of $G$. In particular, $0K$ fixes the average node degree.  $1K$ fixes the node degree distribution, $2K$ fixes the joint node degree distribution, $3K$ fixes the distribution of degree dependent triangles and wedges, and $NK$, where $N=|V|$, fixes the entire graph. It should be noted that \dK fixes $d'K$ for every $d'<d$. According to \cite{Sala2010,sala2011sharing}, $2K$ successfully captures a number of key graph properties, with the exception of clustering.
\end{comment}

%

%

The structure of the rest of the paper is the following. Section \ref{sec:problem} summarizes terminology and the problem statement. Section \ref{sec:related} presents related work. Section \ref{Sec:Estimation from a sample}  discuss the first part of the problem: network sampling, estimation of the 2.5K properties and  post-processing. Section \ref{Sec:2.5K generator} discuss the second part of the problem: a construction algorithm for 2.5K-graphs. Section \ref{sec:evaluation} presents evaluation results on a wide range of fully-known real-life graphs. %
 Section \ref{sec:conclusion} concludes the paper.

\section{Notation and Problem statement}
\label{sec:problem}

We consider an undirected, static graph $G=(V,E)$, with $|V|$~nodes and $|E|$~edges. For the purposes of random walks, we also assume that~$G$ is connected, and aperiodic. For a node $v\in V$, denote by $\deg(v)$ its degree, and by $\mathcal{N}(v)\subset V$ the set of neighbors of~$v$. Let also the number of shared partners between nodes $a$ and $b$ be:  $sp(a,b)=|\mathcal{N}(a)\cap\mathcal{N}(b)|$.

\subsection{Graph Properties of Interest}

{\em Joint Degree Distribution (JDD).} A widely studied property of graphs is the degree distribution. In this paper, we are interested in more information captured by the joint node degree distribution, defined as the number (or frequency) 
of edges connecting nodes of degree $k$ with nodes of degree $l$:

\begin{equation}
\label{eq:JDD}
	\JDD(k,l) = \sum_{a\in V_k}\sum_{b\in V_l} 1_{\{\{a,b\}\in E\}}. 
\end{equation}

In other words, \JDD quantifies a degree-dependent distribution of subgraphs of 2 nodes.

{\em Clustering.} One of the most important topological properties, especially for OSNs, is clustering. The clustering coefficient $c_v$ of a node $v$ captures how close the neighbors of a node are to forming a clique and is typically defined as the ratio of the number of links between the neighbors divided by the maximum number of such links.   If two neighbors of a node are connected, then these three nodes form a triangle, thus leading to an equivalent definition \cite{Watts1998}: 
\begin{equation}\label{eq:c_v}
	c_v = \frac{2 T_v}{\deg(v)(\deg(v)-1)},
\end{equation}
where $T_v$ is the number of triangles using $v$. 
At a slightly coarser granularity, %
the {\em degree-dependent average clustering coefficient} \CCK is defined as 

\begin{equation}\label{eq:c_v_k}
	\bar{c}(k) = \frac{1}{|V_k|}\sum_{v\in V_k} c_v,
\end{equation}
where $V_k$ is a set of nodes of degree $k$. 

Finally,  $\bar{c}$, the clustering coefficient $c_v$ averaged over all nodes in~$G$, is defined as follows: 
{
\begin{equation}\label{eq:c}
	\bar{c} = \frac{1}{N}\sum_v c_v. 
\end{equation}
Note that\ $c_v$  determines both $\bar{c}(k)$ and $\bar{c}$ and $\bar{c}(k)$ determines~$\bar{c}$, because $\bar{c} = \frac{1}{N}\sum_k |V_k|\cdot\bar{c}(k)$. So, \eqn{eq:c}, \eqn{eq:c_v_k} and \eqn{eq:c_v} impose increasingly restrictive constraints on clustering.

\subsection{dk-series} 

 In this paper, we follow and build on the systematic framework of dK-series \cite{Mahadevan2006a}, which characterizes the properties of a graph using series of probability distributions specifying all degree correlations within d-sized subgraphs of a given graph G.   Essentially, dk-series extend the notion of \JDD to any $d$-sized subgraphs. 
To be more concrete:
\begin{itemize}
	\item $0K$ specifies the average node degree.  
	\item $1K$ specifies the node degree distribution.
	\item $2K$ specifies the joint degree distribution (JDD), Eq.(\ref{eq:JDD}) %
	\item $3K$ specifies the degree-dependent distribution of subgraphs of 3 nodes, \ie 
	the number of triangles and wedges connecting nodes of degrees $k, l, m$. 
	\item $NK$, where $N=|V|$, specifies the entire graph.  
\end{itemize}
Clearly,  increasing values of $d$ capture progressively more properties of $G$ at the cost of more complex representation of the probability distribution. %
dK determines $d'K$ for every $d'<d$. The term ``dk-graphs'' refers to all graphs that have the same $d'k$ distributions for $d'=0, 1, ...d$.

\subsection{\em 2.25K and 2.5K graphs} 
According to \cite{Sala2010,sala2011sharing}, $2K$  captures a number of key graph properties, with the exception of clustering, which is inherent in most OSNs. On the other hand, all the metrics of clustering are completely determined by the $3K$ distributions. However, we argue that $3K$ is not practical: it requires prohibitively many samples to be estimated and it is difficult to generate in practice. 
To address these problems, we introduce more practical notions that, intuitively, lie between of $2K$ and $3K$:
\begin{itemize}
\item $2.5K$ specifies \JDD and \CCK: our proposed approach.
\item $2.25K$ specifies \JDD and $\bar{c}$:  a baseline for comparison.
\end{itemize}

\subsection{Problem Statement and Approach}
Our objective is to provide a complete, practical methodology for generating graphs that resembles a real  graph of interest. We use 2.5K as the modeling tool. The problem can be decomposed into two parts:
\begin{itemize}
\item {\em Estimation:} Given a random walk sample of a real graph, estimate ${\JDD}(k,l)$ and ${\bar{c}}(k)$.
\item {\em Graph Generation:} Given desired (and realizable) ${\JDD}(k,l)$, ${\bar{c}}(k)$, construct a graph with those properties. 
\end{itemize} 
The steps of  our approach are summarized  in Fig. \ref{fig:overview}.

\section{Related Work}\label{sec:related}

{\em Network Sampling and Estimation.} Network measurement plays a central role in computer and social network research. In this paper we are mostly interested in the topology of online social networks, such as  Facebook, Twitter and other blogging networks, Linkedin, Instagram (to broadcast pictures), and Epinions. 
Due to the large size of these networks, sampling is typically used to estimate properties of interest. Recent approaches, including our own prior work on sampling OSNs,  used random walks to sample OSNs and estimate nodal attributes and local structural properties \cite{Gjoka2010,Ribeiro2010,Gjoka2011_multigraph_JSAC,Kurant2011_SWRW}.  However,  estimating global structural properties based on sampling remains a challenging problem. 

To characterize the global network structure, model fitting is used.  Handcock et. al.  used the ERGM framework for network inference based on sampled data \cite{handcock2010modeling}. However, ERGM suffers from degeneracy issues and does not scale to even moderate graph sizes of thousands of nodes. Kim et. al.  use Kronecker graphs to fit the network structure of the observed part of the network and then estimate the missing nodes and edges \cite{kim2011network}. Sala et. al.  present an evaluation of model fitting for fully known social graphs and conclude that the dk-series \cite{Mahadevan2006a} is the best model for such task \cite{Sala2010}, surpassing even Kronecker graphs. %

{\em Graph Generation (a.k.a Construction).} Generating random graphs that have some desired properties is an active research area. The complexity of the algorithm and the ability to provide guarantees depend on the desired properties.  For example, 
$1K$ can be generated using the configuration model \cite{molloy1995critical}. $2K$ can be generated using an extension of the configuration model \cite{Mahadevan2006a}.  Although the above approaches may result in multi-edges and self-loops, they can be tweaked to avoid that in both $1K$~\cite{DelGenio2010} and $2K$~\cite{Stanton2011} case. Unfortunately, these construction algorithms do not generalize to $dK, d>2$ and, to the best of our knowledge, no efficient algorithm exists today for generating $3K$ graphs.

However, $1K$ or $2K$ are not sufficient to capture many crucial graph properties, such as (higher than random) clustering, which is inherent in virtually all real-life networks including OSNs \cite{Sala2010,sala2011sharing}. For this reason, the following  algorithms attempt to construct random graphs with some notion clustering. 
\cite{Newman09_RG_with_clustering} extends the configuration model to generate random graphs with a given number of triangles. However, the resulting triangles rarely share common edges,  which results in small values of $c_v$ and prevents us from targeting $\bar{c}$ and $1K$ at the same time. 
 \cite{Bansal2009} targets $1K$ and $\bar{c}$ using an MCMC approach. 
\cite{Serrano2005} proposes a construction algorithm that targets $\bar{c}(k)$ while preserving $1K$;
 they have no control over assortativity, which is actually determined by $\bar{c}(k)$ and $1K$. 
In \cite{Mahadevan2006a}, the authors target $3K$. The approach is to target $3K$ by $2K$-preserving random rewiring.  This approach is unfortunately not practical: we contacted the authors of \cite{Mahadevan2006a} who released only the code for $2K$ construction, which, to the best of our knowledge, is the most advanced application of $dK$-series that is achievable in realistic time frames.

\section{Estimation from a sample}\label{Sec:Estimation from a sample}

As illustrated in \Fig{fig:overview}, our first step is to sample the underlying unknown graph, and to estimate the two properties of 2.5K-graphs, namely 
$\JDD(k,l)$ and \CCK, based on our sample~$S$. 
In this section, we derive such estimators for most common sampling methods: independence sampling (uniform or weighted) or random walk. %
The former is possible if one can sample directly from the userID space, whereas the latter  is the common practice in OSNs via crawling. 

\subsection{Uniform Independence Sampling (UIS)} 
UIS samples the nodes directly from the set~$V$, with replacement, %%
uniformly and independently at random. 

\smallskip
\subsubsection{Estimation of \CCK}
Every triangle $\{a,b,c\}$ contributes exactly count 1 to both $sp(a,b)$ and $sp(a,c)$. 
Therefore
\begin{equation}
T_a \ =\   \frac{1}{2} \sum_{b\in \mathcal{N}(a)} sp(a,b) \ = \ \deg(a) \cdot
\frac
{      \sum_{b\in \mathcal{N}(a)} sp(a,b)}
{2\cdot      \sum_{b\in \mathcal{N}(a)} 1}. 
\end{equation}
 
The latter, seemingly redundant transformation will help us write the estimator. %
Indeed, we are unlikely to cover every node $b\in\mathcal{N}(a)$ in our sample $S$. Instead, they are sampled with  equal probabilities 
and possibly $S.\Count(b)>1$ times. Exploiting the sampled information, we can estimate $T_a$ by 
$$
\est{T}_a \ =\  \deg(a) \ \cdot
\frac
{     \sum_{b\in \mathcal{N}(a)} sp(a,b) \cdot S.\Count(b) }
{2\cdot       \sum_{b\in \mathcal{N}(a)}  S.\Count(b) }.$$
Plugging it into \eqn{eq:c_v}, 
and taking the average across all nodes $a$ of degree $\deg(a)=k$, we obtain
\begin{eqnarray}
\label{eq:c_k_UIS}
\est{\bar{c}}(k)\  
 &=& \frac{1}{k-1} \cdot 
 \frac
 {     \sum_{a\in S_k} \sum_{b\in \mathcal{N}(a)} sp(a,b) \cdot S.\Count(b) }
 {      \sum_{a\in S_k} \sum_{b\in \mathcal{N}(a)}  S.\Count(b)}, \quad{}	
\end{eqnarray}
where $S_k\subset S$ are all the sampled nodes of degree~$k$.

\smallskip
\subsubsection{Estimation of \JDD} 
Let us rewrite \eqn{eq:JDD} as
\begin{eqnarray}
\nonumber \JDD(k,l) & =& |V_k||V_l|\ \cdot \ \frac{    \sum_{a\in V_k}\sum_{b\in V_l} 1_{\{\{a,b\}\in E\}} }{|V_k|\ \cdot\ |V_l|}. 
\end{eqnarray}
The fraction on the right hand side divides the number of \emph{existing} edges between $V_k$ and $V_l$ by the maximal possible number of such edges ($|V_k|\cdot|V_l|$). 
Under UIS, we \emph{observed} $\sum_{a\in S_k}\sum_{b\in S_l} 1_{\{\{a,b\}\in E\}}$ such edges out of the maximal number $|S_k|\cdot|S_l|$ we could possibly observe, leading to the estimator
\begin{eqnarray}
\nonumber \est{\JDD}(k,l) & =& |V_k||V_l|\ 
\frac{   \sum_{a\in S_k}\sum_{b\in S_l} 1_{\{\{a,b\}\in E\}} }{|S_k|\ \cdot\ |S_l|}\\
\label{eq:jdd_uis_induced}    &=& |V_k||V_l|\ 
\frac{    \sum_{a\in S_k}\sum_{b\in \mathcal{N}(a)\cap S_l} S.\Count(b) }{|S_k|\ \cdot\ |S_l|}. \quad{}
\end{eqnarray}
The values $|V_k|$ and $|V_l|$ can be easily estimated as $|V_k|=|S_k|/|S|$ and $|V_l|=|S_l|/|S|$, respectively.

\subsection{Weighted Independence Sampling (WIS)}  
WIS samples the nodes directly from the set~$V$, with replacements, independently at random, but with probabilities proportional to node weights~$\w(v)$. 
For simplicity and compatibility with random walks below, we are interested only in the case where
$	\w(v) = \deg(v)$. 
WIS produces biased (non-uniform) node samples. However, because this bias is known, it can be corrected by an appropriate re-weighting of the measured values, \eg using the Hansen-Hurwitz estimator~\cite{HansenHurwitz1943}, as follows. 

\subsubsection{Estimation of \CCK}

We apply the Hansen-Hurwitz estimator to \eqn{eq:c_k_UIS} by dividing every term related to some node pair by the product of the weights of involved nodes, \ie
\begin{eqnarray}
\label{eq:ck_WIS1} 
\est{\bar{c}}(k) &=&  
\nonumber \frac{1}{k-1}  \cdot
\frac
{   \sum_{a\in S_k} \sum_{b\in \mathcal{N}(a)} \frac{sp(a,b)}{\w(a)\w(b)} \cdot S.\Count(b) }
{  \sum_{a\in S_k} \sum_{b\in \mathcal{N}(a)} \frac{1}{\w(a)\w(b)} \cdot S.\Count(b) } \\
\label{eq:ck_WIS3} &=& 
\frac{1}{k-1}  \cdot
\frac
{   \sum_{a\in S_k} \sum_{b\in \mathcal{N}(a)} \frac{sp(a,b)}{\deg(b)} \cdot S.\Count(b) }
{   \sum_{a\in S_k} \sum_{b\in \mathcal{N}(a)} \frac{1}{\deg(b)} \cdot S.\Count(b) }. \quad {}
\end{eqnarray}
In the last step, we used $\w(v)\!=\!\deg(v)$. The terms $\deg(a)$ cancelled out, because $\deg(a)=k$ for every $a\in S_a$.

\subsubsection{Estimation of \JDD}

Assuming $\w(v)\!=\!\deg(v)$, applying the Hansen-Hurwitz estimator to \eqn{eq:jdd_uis_induced} does not change it (now both $\deg(a)$ and $\deg(b)$ cancel out). 
However, now both $|V_k|$ and $|V_l|$ must be corrected for the biases, as in~\cite{Gjoka2010}
\begin{equation}
|V_k| =  \frac{\sum_{s\in S}\frac{1_{\{\deg(s)=k\}}}{\deg(s)}}{\sum_{s\in S}\frac{1}{\deg(s)}} \qquad 
|V_l| =  \frac{\sum_{s\in S}\frac{1_{\{\deg(s)=l\}}}{\deg(s)}}{\sum_{s\in S}\frac{1}{\deg(s)}}.
\label{eqn:V_L V_l est}
\end{equation}

\subsection{Simple Random Walk (RW)}  
RW selects the next-hop node~$v$ uniformly at random among the neighbors of the current node~$u$.
In a connected and aperiodic graph, the probability of being at the particular node~$v$ converges to the stationary distribution
	$\pi^\RW(v)\ =\ \frac{\deg(v)}{2\cdot|E|}$.

\subsubsection{Estimation of \CCK}

We propose two techniques. %

\paragraph{Induced Edges with safety margin $M$} Generally speaking, in this approach we interpret nodes~$S$ collected by RW as WIS. 
However, because consecutive RW samples are correlated, a straightforward application of \eqn{eq:ck_WIS3} introduces a bias. 
Indeed, RW will observe many more induced edges, defined as edges between any two nodes sampled by RW, than WIS. For example, every step of RW is guaranteed to result in at least one additional induced edge. 
Moreover, these additional induced edges do not follow the same statistical distributions and thus introduce arbitrary biases. 
For this reason, we modify \eqn{eq:ck_WIS3} by ignoring the sample pairs that are closer than margin~$M$  in~$S$:

\begin{eqnarray}
\label{eq:ck_est_M} \est{\bar{c}}(k)\  &=&  \frac
{\displaystyle 
\sum_{\substack{i,j \textrm{ such that }\\ \deg(s_i)=k,\  |j-i|>M,\ \{s_i,s_j\}\in E}} 
\frac{sp(s_i,s_j)}{\deg(s_j)}  
}
{\displaystyle 
\sum_{\substack{i,j \textrm{ such that }\\ \deg(s_i)=k,\  |j-i|>M,\ \{s_i,s_j\}\in E}}
\frac{1}{\deg(s_j)}  
}.\quad{}
\end{eqnarray}
One can check that for $M=0$, the above reduces to \eqn{eq:ck_WIS3}. 
Increasing $M$ makes \eqn{eq:ck_est_M} more robust to RW correlations, at the cost of discarding information. For practical applications, we recommend values $10<M<100$.

\paragraph{Traversed Edges} This technique is based on the observation that edges $S_E$ traversed by RW 
are asymptotically uniform~\cite{Ribeiro2010}, which leads to
\begin{equation}\label{eq:c(k)_RW}
	\hat{c}(k) \ =\  
	\frac{1}{k-1} \cdot 
	\frac
	{\displaystyle \sum_{(u,v)\in S_E} sp(u,v)\cdot \left(1_{\{\deg(u)=k\}}+1_{\{\deg(v)=k\}}\right)}
	{\displaystyle \sum_{(u,v)\in S_E} \left(1_{\{\deg(u)=k\}}+1_{\{\deg(v)=k\}}\right)}.
\end{equation}

\subsubsection{Estimation of \JDD}

Similarly to the estimation of $c(k)$, we propose two techniques.

\paragraph{Induced Edges with safety margin $M$}

The ``safety margin'' $M$ trick we used to correct the clustering coefficient estimator can be also applied here. To this end, we modify \eqn{eq:jdd_uis_induced} by ignoring the sample pairs that are closer than margin~$M$ in $S$, which results in
\begin{eqnarray}
\est{\JDD}(k,l)  &=&  |V_k||V_l| \cdot \frac
{\displaystyle 
\sum_{\substack{i,j \textrm{ such that }|j-i|>M,\\ \deg(s_i)=k,\ deg(s_j)=l}  }
1_{\{s_i,s_j\}\in E}
}
{\displaystyle 
\sum_{\substack{i,j \textrm{ such that }|j-i|>M,\\ \deg(s_i)=k,\ deg(s_j)=l}} 
1
},\qquad{}
\end{eqnarray}
where $|V_k|$ and $|V_l|$ are calculated as in \eqn{eqn:V_L V_l est}.

\paragraph{Traversed Edges} 
As before, an alternative approach is to interpret edges $S_E$ traversed by RW as asymptotically uniform. Now, we just check what fraction of edges traversed by RW are between nodes of degree $k$ and $l$, and then we inflate this fraction by $|E|$, as follows:
\begin{equation}\label{eq:JDD_RW}
	\est{\JDD}(k,l) \ =\   |E| \cdot 
	\frac{      \sum_{(u,v)\in S_E}  \left(1_{\{\deg(u)=k,\deg(v)=l\}}\right)}
	{     |S_E|}.
\end{equation}

\subsubsection{Hybrid Estimators}
Under RW, we described two general estimation techniques, Traversed Edges~(TE) and Induced Edges~(IE). 
They lead to two different estimators. Which one should we choose? The answer depends on the graph size and structure, on what we are trying to estimate, and on the sample size. 
For example, TE visits one edge per iteration, so its collected and exploitable information grows linearly with the sample size~$n$. 
In contrast, while for small $n$ IE may include very few edges, it quickly catches up for larger~$n$. 
So the first obvious hint is to compare the number of traversed and induced edges. 
Moreover, while TE samples edges uniformly, IE will be more likely to cover edges connecting nodes with high degree. Consequently, we expect TE to perform better when estimating values related to low degree nodes. 

In order to combine the advantages of both TE and IE, we use the estimate of TE  for small degrees and the estimate of IE for large degrees. In order to switch between TE and IE, we compare the actual degree(s) to a threshold, which we choose to be the average node degree.  These \emph{hybrid estimators} are the ones we use in our approach: 
\begin{eqnarray}
\label{eq:hybrid}
  \est{\bar{c}}^\sss{hybrid}(k) &=& \left\{ 
  \begin{array}{l l}
    \est{\bar{c}}^\sss{TE}(k)  & \quad \text{ if  $k\!<\!\bar{k}$}\\
    \est{\bar{c}}^\sss{IE}(k) & \quad \text{ otherwise. }\\
  \end{array} \right.\\
  \est{\JDD}^\sss{hybrid}(k,l) &=& \left\{ 
  \begin{array}{l l}
    \est{\JDD}^\sss{TE}(k,l)  & \quad \text{ if  $k\!+\!l\!<\!2\bar{k}$}\\
    \est{\JDD}^\sss{IE}(k,l) & \quad \text{ otherwise. }\\
  \end{array} \right.
\end{eqnarray}

\subsection{Postprocessing of Estimated Parameters}

\subsubsection{Smoothing}

The estimation of the joint degree distribution from node samples produces a considerable number of high frequency elements in the JDD 2-dimensional matrix. For example, degree pair entries that involve one low degree node are overestimated since low degree nodes have lower visiting probability in random walks.
For that reason, we apply Gaussian kernel smoothing to the measured matrix to reduce the amplitude of such discrete elements. We select the kernel bandwidth using Scott's rule of thumb \cite{scott1992multivariate}. 

\subsubsection{Realizable JDD}

There is no guarantee that there exists a simple graph that has the joint degree distribution estimated in section \ref{Sec:Estimation from a sample}.  
 According to \cite{stanton2011sampling}, the necessary and sufficient conditions that make a joint degree distribution realizable  are the following: (i) $\JDD(k,l) \in \mathbb{Z} $; (ii) $deg(k) \in \mathbb{Z} $; (iii) $\JDD(k,l) \leq deg(k)*deg(l),~~k \neq l $; (iv) $ \JDD(k,k) \leq \binom{deg(k)}{2},~~k=l $; (v) $\JDD(k,k) = 2*f_{k},~e_{k} \in \mathbb{Z} $, where  $deg(k)$ represents the number of nodes of degree $k$. 

Conditions (i) and (ii) state that the degree and joint degree distributions must be integer. To that end, we had to stochastically round estimated values of the JDD matrix. Conditions (iii) and (iv) state that the number of edges between nodes of degree $k$ and $l$ cannot exceed the maximum possible number of edges, given the number of nodes $D(k)$  and $D(l)$. Last, (v) states that the number of edges between nodes of the same degree has to be even.

It is imperative that the estimated JDD matrix fed into the 2.5K generator is realizable. Otherwise, the constructed graph might contain nodes with degrees that were not sampled  from the graph.  That creates a problem during the phase in which we target the measured $\bar{c}(k)$, since the latter will only contain degrees that were sampled from the graph. For example, assume that our estimators yield $\sum_{k} JDD(10,k) = 46$, \ie there are 46 edges connected to nodes of degree 10. If we construct a 2K graph using this JDD matrix, we will get 4 nodes of degree 10 and 1 node of degree 6. There is a problem if the estimators also yield $\sum_{k} JDD(6,k) = 0$, which means that there are no nodes of degree $6$.

To address this problem, we designed an algorithm that slightly modifies the JDD matrix to make it realizable. We formulated the problem as an optimization problem (to minimize the error between the estimated and the modified JDD subject to realizability constraints) and we also developed an efficient algorithm that provably achieves a realizable JDD. In the above example, our algorithm stochastically  chooses whether to keep 5 or 4 nodes of degree 10 and then  adds accordingly 4  or removes 6 edges in the JDD matrix, while satisfying the above conditions. Due to lack of space we omit the details and defer the full description to  the source code at \cite{2.5K_source_code}.

In summary, at the end of this step, we have slightly modified the estimated JDD to make it realizable and ready to be fed to the generation algorithm in the next step.  We verified that the changes were indeed minimal: in all simulations, postprocessing  changed no more than 3\% of the edges. It is worth noting that postprocessing is not necessary for the estimated $\bar{c}(k)$, since the construction algorithm achieves JDD exactly but clustering only approximately.

\section{Generating a $2.5K$ graph}
\label{Sec:2.5K generator}

In this section, we design an algorithm that takes as input the two target properties estimated as in the previous section, \ie the  target joint node degree distribution, $\target{\JDD}(k,l)$ and the target degree-dependent average clustering, $\target{\bar{c}}(k)$, and constructs a 2.5K-graph with $N$ nodes and the target properties.

\subsection{Unsuccessful attempts and lessons learned}

\subsubsection{MCMC}\label{subsec:MCMC}
The authors of \cite{Mahadevan2006a} apply an extension of the configuration model~\cite{molloy1995critical} to generate a graph that exactly satisfies $\target{\JDD}(k,l)$. Starting from such a graph, one can perform $2K$-preserving double-edge swaps to target the clustering $\target{\bar{c}}(k)$, as follows:
\line(1,0){250}\\
MCMC\\
 \DO \\
  \SP randomly select edges $(u,v)$ and $(x,y)$ such that $k_u=k_x$\\
  \SP rewire these edges into $(u,y)$ and $(x,v)$\\
  \SP \IF $\sum_k|\target{\bar{c}}(k)-\bar{c}(k)|$ has increased \THEN\\
  \SP\SP undo the rewiring\\
\line(1,0){250}

Although \cite{Mahadevan2006a} proposed this method to target the entire $3K$, we found it impractical already for its relaxed version $\bar{c}(k)$: very soon after creating the first triangles, there is very small probability that edge swap
brings us closer to the target. Consequently, this MCMC approach takes forever in practice.

\subsubsection{Improved MCMC}
We tried to address this problem by selecting the two candidates for a swap in a smarter way, e.g., by favoring edges with fewer triangles attached. The rationale was that after deleting these edges few triangles are destroyed. Although this improved over the naive MCMC, still we still faced scalability problems. %

\subsection{Our $2.5K$ generator}

Our key insight is that, for the same reason why it is difficult to create triangles with double-edge-swaps, it is easy to destroy the existing triangles. This suggests starting the MCMC from triangle-rich $2K$ graph rather than with a regular, triangle-poor $2K$ graph,  as that used in \cite{Mahadevan2006a}.
If the starting graph overshoots the target $\target{\bar{c}}(k)$, the job of the MCMC phase is to destroy triangles rather than creating new ones, which is much faster.

{\em Step 1.} In order to create a triangle-rich graph with a given $\target{\JDD}(k,l)$, we initially follow  the two initial  steps of~\cite{molloy1995critical} and~\cite{Mahadevan2006a}: we create a set of nodes $V$, where $|V|=N$, and we assign \emph{target degree} $\target{k}_v$ to every node $v\in V$ such that the target 1K distribution (fully defined by $\target{\JDD}(k,l)$) is satisfied. Next, we apply the following algorithm.
\line(1,0){250}\\
{\em Step 2.}  Greedily create local edges:\\ 
\REQUIRE \JDD(k,l)\\
 \FOR $v\in V$ \DO $r_v=rand(0,1)$ \\
 $dist(u,v)\ =\ \min(|r_v-r_u|, 1-|r_v-r_u|)$\\
 $E'\gets$ a list of all possible node pairs $\{u,v\}$\\
 sort $E'$ according to $dist(u,v)$\\
 $E=\emptyset$\\
 \FORALL $\{u,v\}\in E'$ \DO\\
 \SP	\IF  {\small $\JDD(k_u,k_v)\smaller\target{\JDD}(k_u,k_v)$  \small \AND $k_u\smaller\target{k}_u$ \AND $k_v\smaller\target{k}_v$ \DO}\\
 \SP\SP  $E\gets E\cup \{u,v\}$\\
 \line(1,0){250}\\ %%%%%%%%%%%
The above algorithm first assigns to every node $v$ a coordinate $r_v$ randomly selected from interval $(0,1)$. 
Then, it creates a set $E'$ of all possible node pairs sorted by increasing distance in this one-dimensional coordinate system. Finally, it goes through all pairs in $E'$ and creates an edge if the target values $\target{\JDD}$ and $\target{k}$ are not exceeded. This construction ensures that  the created edges tend to be local (\ie with small $dist(u,v)$), which leads to many triangles. (Notice that, throughout the algorithm execution, the target degree~$\target{k}_v$ and joint degree~$\target{\JDD}(k,l)$ remain unchanged, while the current node degree~$k_v$ and joint node degree $\JDD(k,l)$ may change with every added/modified edge. In the beginning, we have $k_v=0$ and $\JDD(k,l)=0$, because there are no edges in the graph yet.) 
 If we reach the target values $\target{\JDD}$ and $\target{k}$ at the end of this step, we are done. If the target values are not reached, we are in the situation depicted in \Fig{fig:trick_all}(a) and we need to throw some more edges as follows.

{\em Step 3.} We iteratively apply the transformations described in \Fig{fig:trick_all}, until our graph satisfies precisely the target $\target{\JDD}(k,l)$ (thus $\target{k}_v$ as well). 
In our simulations, we saw that the number of created triangles at the end of this step, dramatically exceeds the targeted ones, to achieve~$\target{\bar{c}}(k)$, as shown in \Fig{fig:neworleans_targetproperties}. 

{\em Step 4.} Finally, we apply the 2K-preserving, $\target{\bar{c}}(k)$-targeting double-edge swaps MCMC, described in \Sec{subsec:MCMC}. 

We release an implementation of our $2.5K$ generator at \cite{2.5K_source_code}.

\subsection{Guarantees}
Our algorithm is a heuristic, in the sense that it does not currently come with provable guarantees. However, it is worth noting that in {\em all} our simulations we were able to construct graph instances that had always the {\em exact} target JDD and {\em approximately} the target clustering (closer to the target and order of magnitudes faster than  prior approaches). 
Theoretical guarantees for the achieved properties  and a characterization of the variability of the constructed graphs are possible directions for future work.

\begin{figure}[t!]
\psfrag{A}[l][c][1]{a)} 
\psfrag{B}[l][c][1]{b)}
\psfrag{C}[l][c][1]{c)}
\psfrag{D}[l][c][1]{d)}
\psfrag{E}[l][c][1]{e)}
\psfrag{F}[l][c][1]{f)}
\psfrag{G}[l][c][1]{g)}
\psfrag{H}[l][c][1]{h)}
\psfrag{a}[l][c][1]{$a$} 
\psfrag{b}[l][c][1]{$b$} 
\psfrag{c}[l][c][1]{$c$} 
\psfrag{d}[l][c][1]{$d$} 
\psfrag{L2}[l][c][0.70]{Nodes with unreached target degree, \ie with $k_v<\target{k}_v$ (here $k_v=\target{k}_v-1$).}
\psfrag{L0}[l][c][0.70]{Nodes with $k_v=\target{k}_v$. }
\centering
\includegraphics[width=0.49\textwidth]{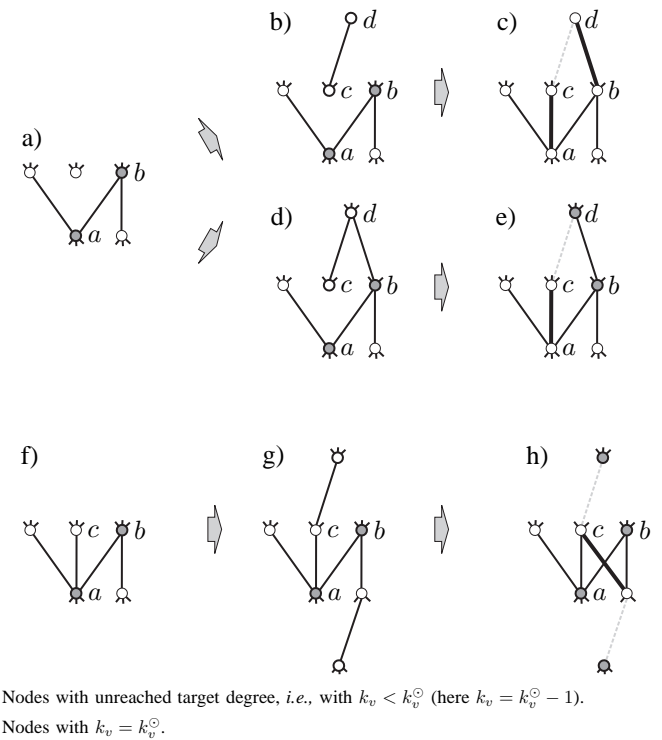}
\caption{{\bf  Illustration of Step 3.} Throwing remaining edges. Several cases:  
\quad \textbf{A)}~Nodes with target degree $\target{k}$ (top) and $\target{l}$ (bottom). 
Assume that current $\JDD(k,l)=\target{\JDD}(k,l)-1$, so we still have one edge left to throw. 
The only two nodes with unreached target degree are $a$ and $b$, but an edge $(a,b)$ already exists. 
\quad \textbf{B)}~In that case, we find nodes $c$ and $d$ such that:
(i) $\target{k}_c = \target{k}$,  
(ii)~$(a,c)$ does not exist, 
(iii)~$(c,d)$ exists, and
(iv)~$(d,b)$ does not exist. 
\quad \textbf{C)}~Create $(a,c)$, and change $(c,d)$ into $(b,d)$. As a result, $\JDD(k,l) = \target{\JDD}(k,l)$, $\JDD()$ of all other degree pairs remain untouched, $k_c$ and $k_d$ remain the same, and $k_a=\target{k}_a$ and $k_b=\target{k}_b$. 
\quad \textbf{D,E)}~If $(b,d)$ exists for every $d$ (rarely happens in practice), follow \textbf{(b)} and \textbf{(c)} without creating $(b,d)$. Now, the problem is moved to another pair of degrees. 
\quad \textbf{F)}~Finally, it is possible that edge $(a,c)$ exists for all candidates~$c$. 
\quad \textbf{G,H)}~In that case, add an edge between two nodes that reached the target degree (here $c$ and $e$) and delete one edge of $c$ and one of $e$. As a result, we have two pairs of nodes to deal with. 
}
\vspace{-16pt}
\label{fig:trick_all}
\end{figure}

\section{Performance Evaluation} 
\label{sec:evaluation}

In this section, we evaluate the performance of our approach.  First, we evaluate the efficiency of the estimators of the 2.5K parameters. Then, we show that our 2.5K generator is orders of magnitude faster than state-of-the-art approaches. We also show that the generated graphs are very close to the original ones with regards to a number of graph properties (beyond JDD and clustering, which are met by construction).

\subsection{Simulation Setup}

\begin{table}[h]
\vspace{-5pt}
  \centering
  {
\begin{tabular}{|r@{}|r|r|r|r|r|}
\hline
    Dataset          & $|V|$   & $|E|$ & $k_V$ & $\displaystyle{\sum_{v\in V}^{} T_v}$ & $\bar{c}$ \\
\hline
FB: UCSD~\cite{Traud2011}&      14\,948 &     443\,221 &  59.30 &  7\,995\,471  & 0.227\\
FB: Harvard~\cite{Traud2011} &  15\,126 &    824\,617  & 109.03 & 24\,848\,793   & 0.212\\
FB: New Orl.~\cite{Viswanath2009} &      63\,392 &     816\,884 &  25.77 &  10\,504\,548  & 0.222\\
soc-Epinions~\cite{WWW_SNAP_Graph_Library}&    75\,877 &     405\,737 &  10.69 &  4\,873\,260  & 0.138\\
email-Enron~\cite{WWW_SNAP_Graph_Library}&    36\,692 &     183\,831 &  10.02 &  2\,181\,132  & 0.497\\
CAIDA AS ~\cite{WWW_SNAP_Graph_Library}&      26\,475 &     53\,377 &  4.03 &  109\,086  & 0.208\\
\hline
\end{tabular}}
\vspace{-2pt}
\caption{Empirical topologies used in Sec.~\ref{sec:evaluation} %
}
\vspace{-10pt}
\label{tab:Topologies}
\end{table}

{\em Data Sets.} \Tab{tab:Topologies} lists the real topologies that we use in our evaluation.  The list includes online social networks, email communication graphs and autonomous systems graphs. The average degree varies from 4 to 109 and clustering varies from 0.14 to 0.50. We treat all topologies as undirected graphs. 

{\em Comparison of Graph Properties.} We measure the difference between two discrete distributions using Normalized Mean Absolute  Error (NMAE) defined as:
$\label{eq:NMAE}
NMAE(\est{\vec{x}},\vec{x}) = \frac{ \sum(|\est{x}_i-x_i|)  }{\sum x_i},
$
where~$\vec{x}$ and $\est{\vec{x}}$ are the vectors that correspond to the real and estimated discrete distributions. $NMAE$ returns the percentage of error, averaged over every point in the discrete distribution.

\subsection{2.5K Estimation and Postprocessing}

\begin{figure*}[t]
\centering
\subfigure[ $\est{c}(k)$ for samples of length 3\%.]{
 \includegraphics[width=0.33\textwidth]{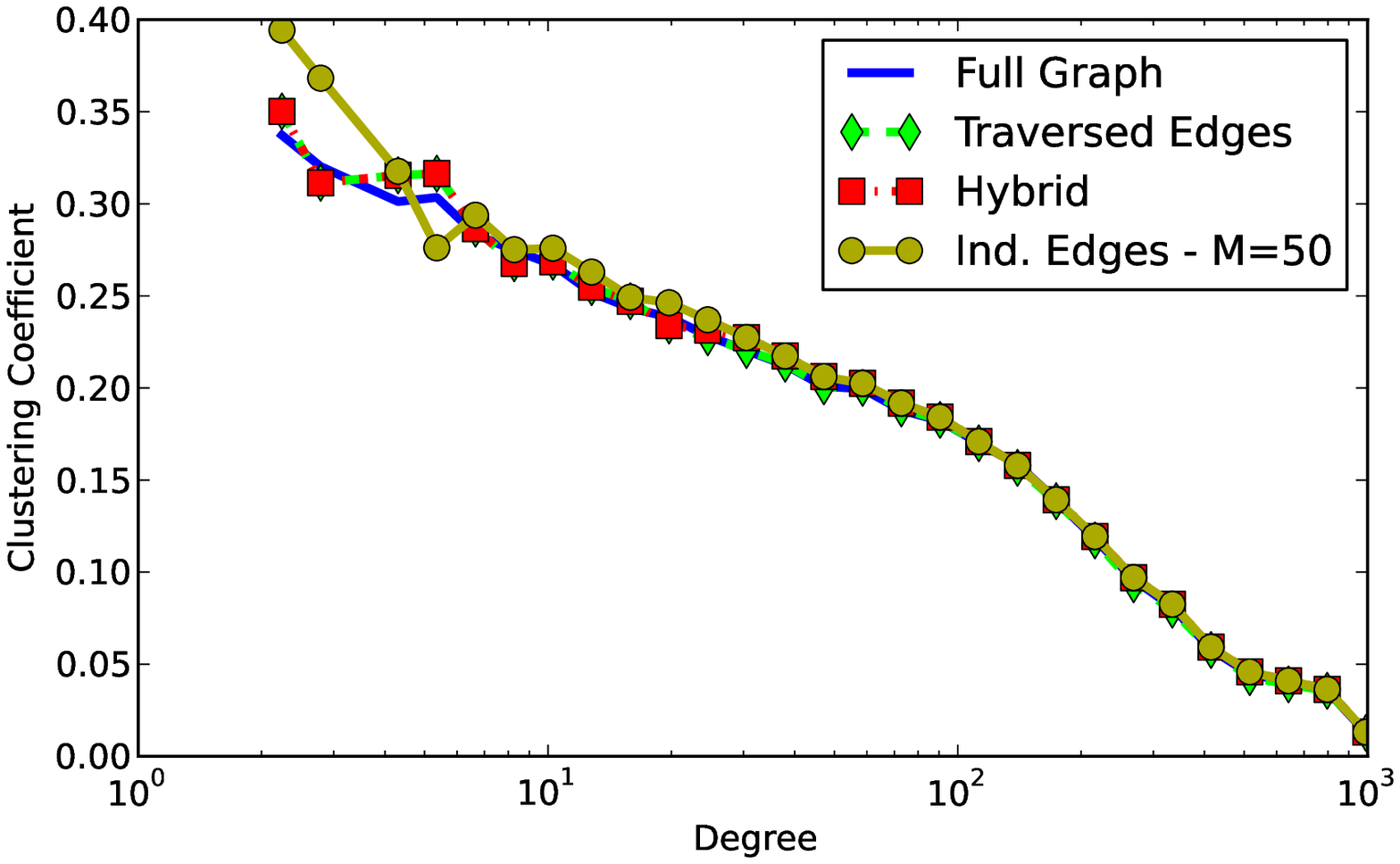} 
 \label{estimation_a}} \hspace{-12pt}
\subfigure[ $\est{c}(k)$ for sample length 1\%-40\%]{
 \includegraphics[width=0.33\textwidth]{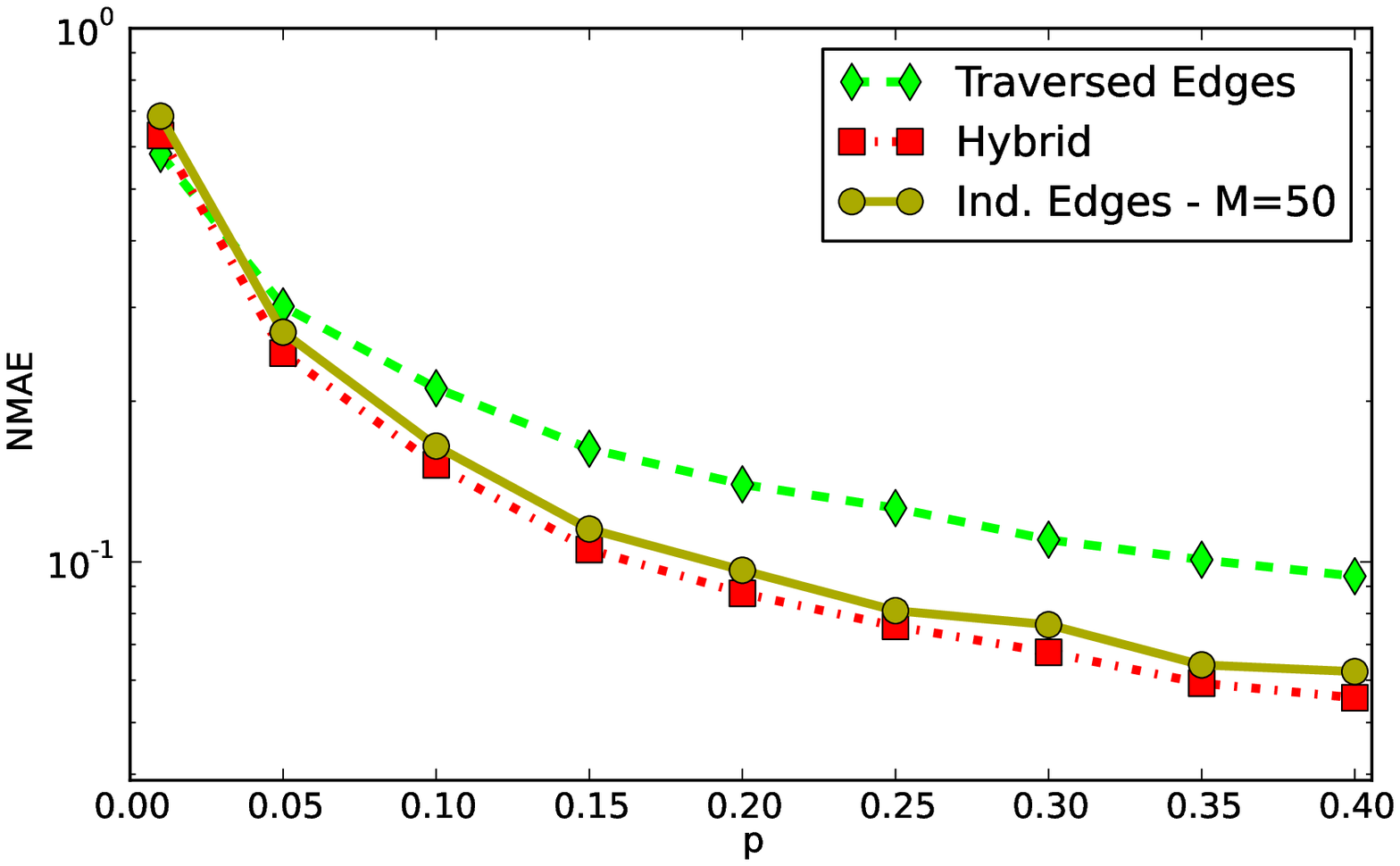}
 \label{estimation_b}} \hspace{-12pt}
\subfigure[ $\est{JDD}(k,l)$ for sample length 1\%-40\%]{
 \includegraphics[width=0.325\textwidth]{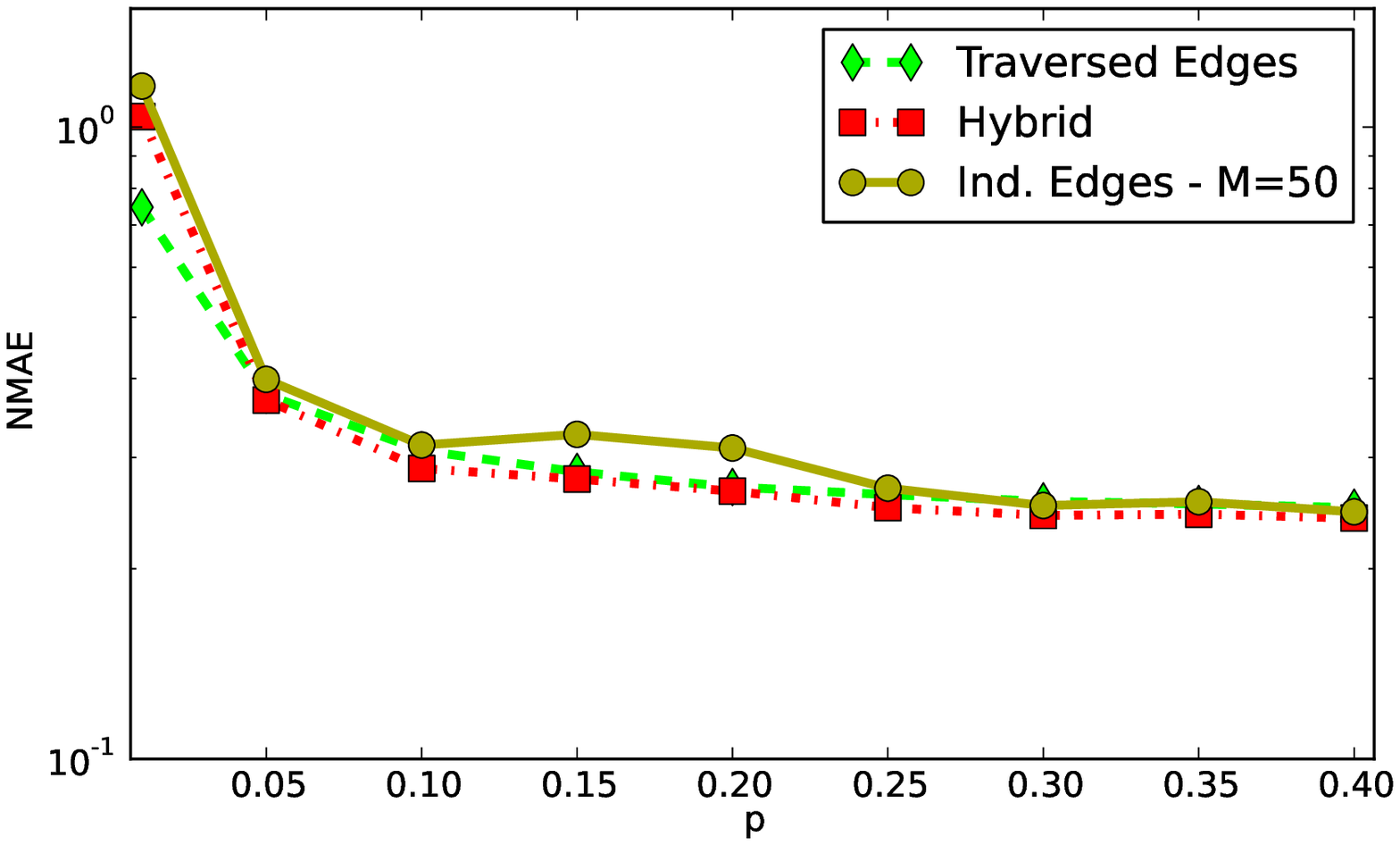}
 \label{estimation_c}} \hspace{-12pt}
\vspace{-7pt}
\caption{{\em Facebook New Orleans} Estimation of clustering $\est{c}(k)$ and $\est{JDD}(k,l)$ with smoothing. The results are aggregated over 100 samples. }
\vspace{-11pt}
\label{fig:neworleans_estimation}
\end{figure*}

{\em Estimation.} In this part, we test our estimators of $JDD(k,l)$ and $\bar{c}(k)$, developed in \Sec{Sec:Estimation from a sample}.  Previously, we introduced two techniques to estimate $JDD(k,l)$ and $\bar{c}(k)$. We argued that Traversed Edges is better for very small degrees.
\Fig{estimation_a} demonstrates this point in the Facebook New Orleans network. For a sample length of 3\%, Traversed Edges better estimates the clustering coefficient for degrees $k<30$. 
\Fig{fig:neworleans_estimation}(b,c) show that the Hybrid estimator, defined in \eqn{eq:hybrid},  outperforms the two base estimators for sample length 1\%-40\% in the estimation of $\bar{c}(k)$ and $JDD(k,l)$ . In the rest of the experiments, we always use the Hybrid estimator.

\begin{figure*}[t]
\centering
\subfigure[Full graph]{\includegraphics[width=0.30\textwidth]{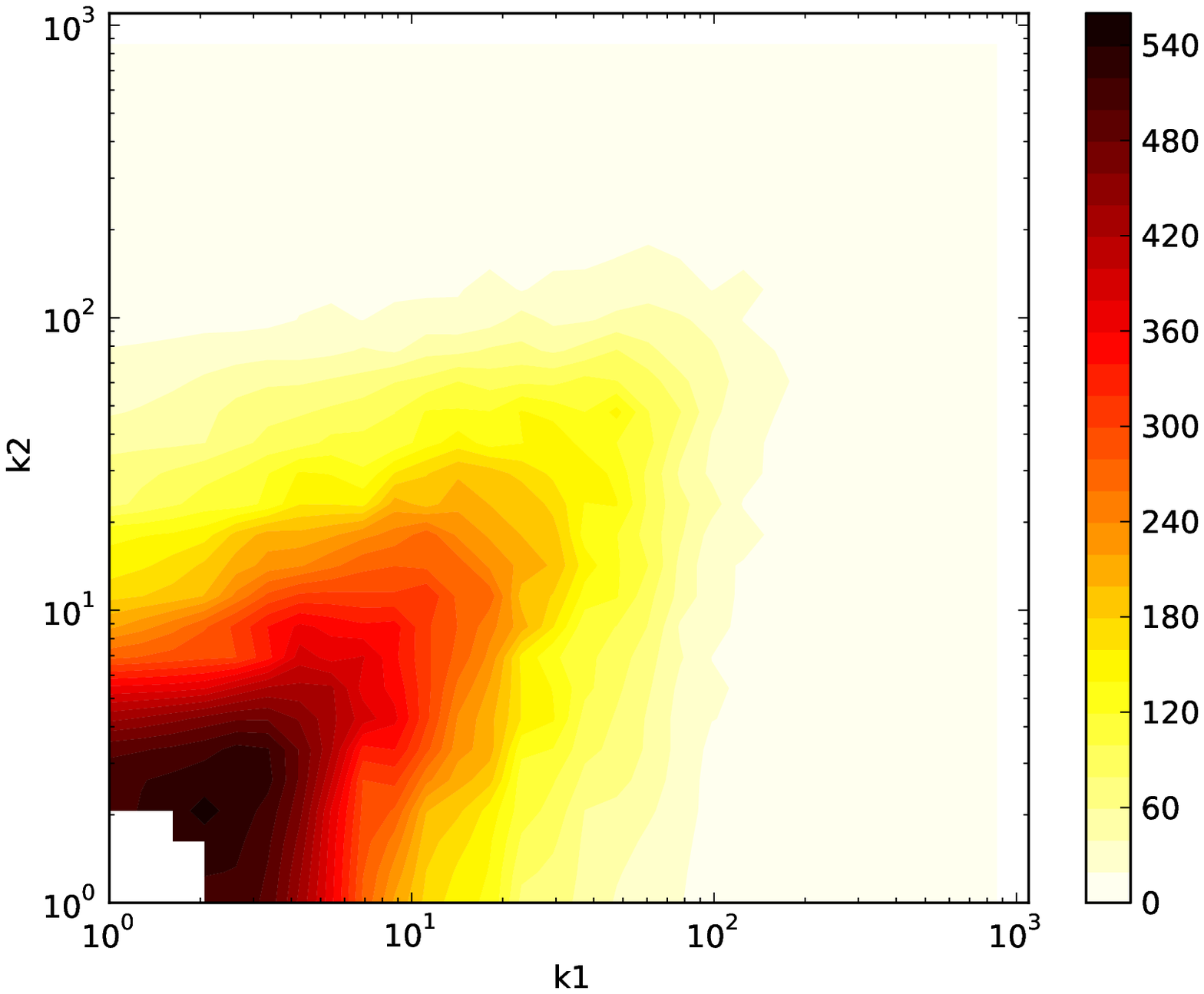} \label{smoothing_a}}
\subfigure[20\% sample w/out smoothing (NMAE:0.52)]{\includegraphics[width=0.30\textwidth]{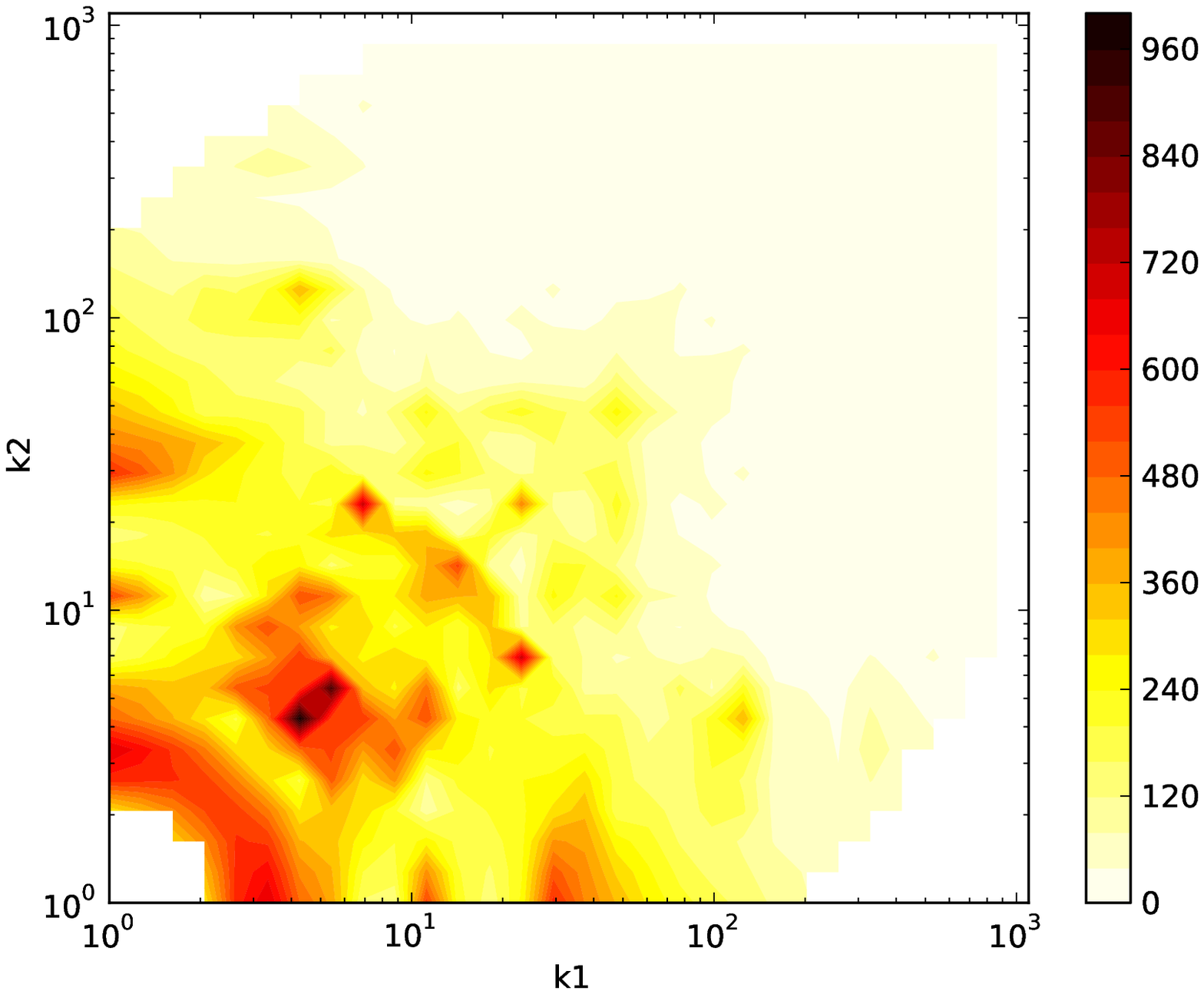}\label{smoothing_b}}
\subfigure[20\% sample with smoothing (NMAE:0.25)]{\includegraphics[width=0.30\textwidth]{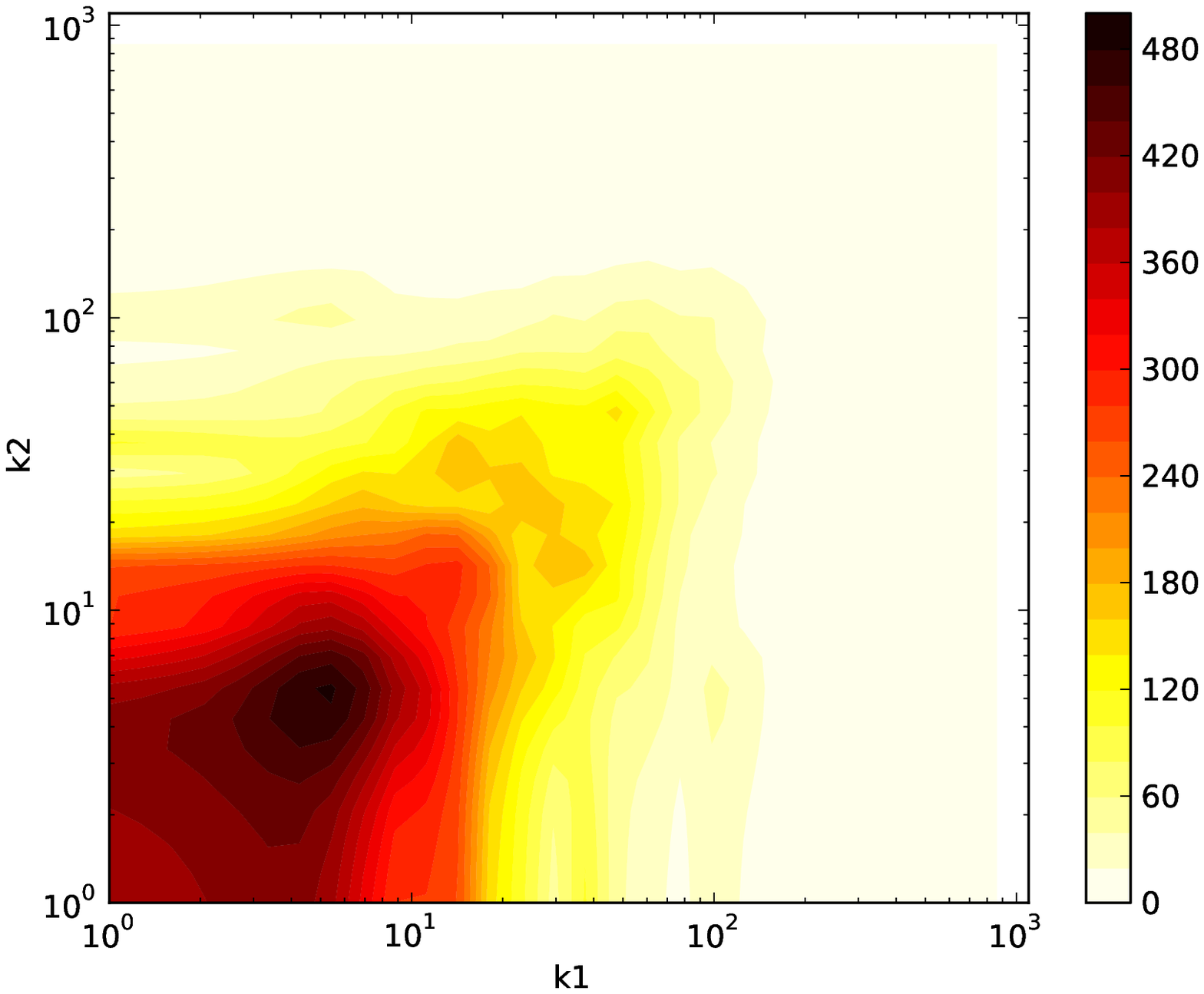}\label{smoothing_c}}
\vspace{-7pt}
 \caption{{\em Facebook New Orleans} Estimation of Joint Degree Distribution $\est{\JDD}(k,l)$.  The effect of smoothing.}
\vspace{-14pt}
\label{fig:nkk_smoothing_neworleans}
\end{figure*}

{\em Postprocessing.}  After the estimation of $JDD(k,l)$ we 
smooth the high frequency elements of the matrix and ensure realizability during construction. \Fig{fig:nkk_smoothing_neworleans} shows the effect of smoothing on the Facebook New Orleans network. \Fig{smoothing_b} and \Fig{smoothing_c} are the non-smoothed and smoothed versions  of a 20\% sample length random walk. The smoothed version has considerably smaller error (NMAE 0.25 vs 0.52) and its highest frequency element is closer to the full graph (480 vs 960) . \Fig{fig:nkk_smoothing_neworleans} also provides a visual validation of the estimation result. Last, the modification of the JDD matrix to make it realizable results in a small number of edge changes in the matrix, typically between 1\%-5\%. Due to lack of space, we omit additional results.

\subsection{$2.5K$ Graph Generation}

\subsubsection{Speed of Generation}

\begin{table}[t!]
  \centering
  {
\begin{tabular}{|r@{}|r|r|r|r|}
\hline
    Dataset                                      & 2K-T +      &   2K-T +  &  2K +       &   2K + \\
                                                 & Imp. MCMC   &   MCMC    &  Imp. MCMC  &  MCMC  \\
\hline
FB: UCSD~\cite{Traud2011}                        &  568   &  1\,742    &   24\,800  &  177\,533 \\
FB: Harvard~\cite{Traud2011}                     &   1\,182    &  2\,880    &  50516  & 387\,506  \\
FB: New Orl.~\cite{Viswanath2009}                                 &  1\,463   &  5\,450   &  118\,711  &  381\,397  \\
soc-Epinions~\cite{WWW_SNAP_Graph_Library}       &  888   &  1\,080    &  3\,342  &   8\,958\\
email-Enron~\cite{WWW_SNAP_Graph_Library}        &  4\,279     &  14\,393  &  66\,766  &  196\,202 \\
CAIDA AS ~\cite{WWW_SNAP_Graph_Library}          & 121   &  141   &  131  &   168 \\
\hline
\end{tabular}}
\vspace{-2pt}
  \caption{Graph generation time in seconds.}
 \vspace{-20pt}
  \label{tab:Runtime}
\end{table}

To better understand the gains, we evaluate separately two parts of the 2.5K generator:  the first part  constructs a graph with an exact JDD and the second part  approximately achieves \CCK. In the first part, we compare: (i) a baseline - the algorithm from \cite{Mahadevan2006a}, simply referred to as {2K}; and (ii) our algorithm (steps 1,2,3 in Section V.B),  which we call {\em 2K-T} because it constructs an exact JDD but with a large number of triangles. In the second part, we compare the two options mentioned in Section V.A: (i) MCMC (ii) and Improved MCMC. We ran simulations for all four possible combinations of the two parts to achieve 2.5K on the datasets of \Tab{tab:Topologies}.   
Simulations were performed on an AMD Opteron machine clocked at 3.2 Ghz.  We set as the stopping condition for the second part to:  $NMAE<2\%$. 

\begin{figure}[ht!]
\centering
\vspace{-2pt}
 \subfigure[$NMAE$ and Average Clustering Coefficient in time.]{
\label{fig:neworleans_speed}
\includegraphics[scale=0.36, angle=0]{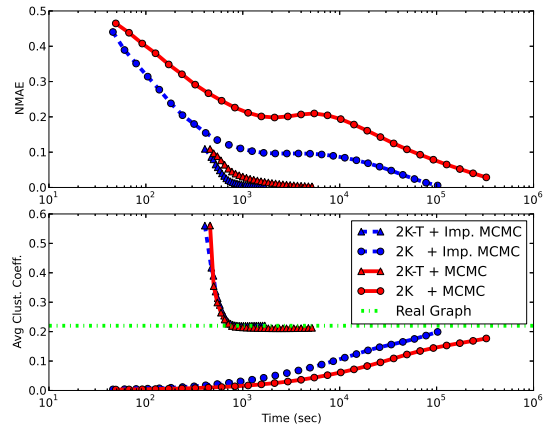}
 }
\vspace{-4pt}
\subfigure[Degree-dependent Average Clustering Coefficient.]{
\label{fig:neworleans_targetproperties}
\includegraphics[scale=0.36, angle=0]{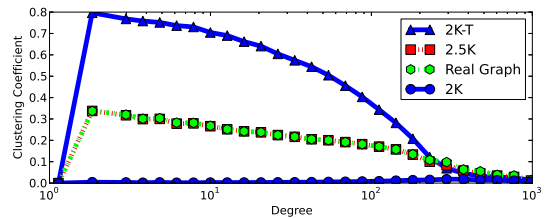}
}
\vspace{-2pt}
\caption{{\em Facebook New Orleans} Speed of 2.5K generation}
\vspace{-16pt}
\end{figure}

We present the simulation results in \Tab{tab:Runtime}. The best performing combination, and thus our proposed method, is  2K-T+Improved MCMC.
It achieves up to 300 times better performance than 2K+MCMC. The speedup we obtain can be decomposed in two parts: 2-6 times because of improved MCMC and up to 50 times because of the 2K-T construction.

We further demonstrate this speedup in a simulation of the New Orleans Facebook network. Fig~\ref{fig:neworleans_speed} shows NMAE error and average clustering as a function of simulation time. The 2K and 2K-T construction time is $\sim$ 40 and 400 sec, respectively. Despite this head start in the first part, the 2K versions take between $118K-381K$ sec whereas the 2K-T versions take between $1K-5K$ sec to target $\bar{c}(k)$, depending on the version of MCMC used. This huge difference in running time is due to the large number of triangles created by our 2K-T construction, as shown in \Fig{fig:neworleans_targetproperties}. This confirms our intuition that, when targeting $\bar{c}(k)$, the task of destroying triangles  is much easier than creating new ones.

We observed that our 2.5K generator (2K-T+Improved MCMC) yields larger performance gains in graphs that have a high number of triangles per node on average. Social graphs and other human communication graphs fall into that category. In contrast, the dataset CAIDA AS is a autonomous system that has low number of triangles/node, even though it has a relatively high global clustering coefficient $\bar{c}$. In this dataset, we observe from \Tab{tab:Runtime} that our $2.5K$ generator performs similarly to $2K+MCMC$ in terms of construction time.

\begin{table}[t]
\scriptsize
  \centering
  {
\begin{tabular}{|@{}l@{}|@{}r@{}|r@{}|r@{}|r@{}|r@{}|r@{}|r@{}|r@{}|@{}r@{}|@{}r@{}|}
\hline
     Dataset                    & Graph            &            \multicolumn{9}{c|}{Norm. Mean Abs. Error. Comparison with Real.}  \\  \cline{3-11} 
                               &  Generation        &              \multicolumn{9}{c|}{Graph properties}   \\   \cline{3-11}
                               &                   &    DD   & Knn   &  JDD  &  CC      &  ESP     & Sh.P.    & Cliq. &  \xspace Cycl. & \xspace Spect.  \\ %
\hline
\multirow{7}{*}{UCSD}         &  2K	           &    0    &  0    &    0  &   0.87   &  1.26  &  0.19 &   1.29  &    0.51     &   0.52   \\ %
	   	               &  2.25K            &    0    &  0    &    0  &   0.22   &  0.58  &  0.06 &   4.54  &  0.21   &   0.13  \\ %
                               &  2.5K             &    0    &  0    &    0  &   0.02  &  0.44  &  0.03 &   3.15  &  0.13   &   0.10  \\ %
			       &  20\% samp.+\xspace\xspace2K &    0.15 &  0.11 &  0.37 &   0.88   &  1.23  &  0.13 &   1.25  &  0.53   &   0.52       \\ %
			       & 20\% samp.+2.5K &  0.15 &  0.11 &  0.37 &   0.14   &  0.44  &  0.18 &   3.4   &  0.14   &   0.10  \\ %
			       &  30\% samp.+\xspace\xspace2K   &  0.16 &  0.08 &  0.35 &   0.88   &  1.26  &  0.09 &   1.27  &  0.53   &   0.52       \\ %
			       &  30\% samp.+2.5K &  0.16 &  0.08 &  0.35 &   0.11   &  0.45  &  0.10 &   3.92  &  0.13   &   0.11   \\ %
\hline \hline
\multirow{7}{*}{Harvard}   &  2K	           &    0    &  0    &   0  &   0.75   &  1.05   & 0.23  &   1.10    &  0.40   &  0.56       \\ %
	   	               &  2.25K            &    0    &  0    &   0  &   0.26   &  0.61  &  0.07  &   1.43    &  0.25 &   0.25  \\ %
                               &  2.5K             &    0    &  0    &   0  &   0.02  &  0.37  &  0.10  &   1.21    &  0.17  &   0.12    \\ %
			       &  20\%  samp.+\xspace\xspace2K &   0.28  &  0.14 &  0.49&   0.80   &  1.09   & 0.09  &  1.08  &  0.43  &   0.57       \\ %
			       &  20\% samp.+2.5K & 0.28  &  0.14 &  0.49&   0.17   &  0.41  &  0.28  &   1.26    &  0.18  &  0.13         \\ %
			       &  30\%  samp.+\xspace\xspace2K &   0.18  &  0.09 &  0.47&   0.78   &  1.05  &  0.19  &  1.10  &  0.43  &  0.57      \\ %
			       &  30\% samp.+2.5K & 0.18  &  0.09 &  0.47&   0.11   &  0.38  &  0.08  &   1.22   &  0.19  &  0.12          \\ %
\hline \hline
\multirow{7}{*}{New Orl.}     &  2K	           	        &    0    &  0    &   0  &   0.92   &  1.3   &  0.35   &  1.20   &  0.68  &    0.54    \\ %
	   	               &  2.25K                         &    0    &  0    &   0  &   0.19   &  0.45  &  0.15   &  1.55  &  0.21  &    0.06   \\% &   0.025     \\
                               &  2.5K                          &    0    &  0    &   0  &   0.02   &  0.33  &  0.05   &  1.50  &  0.18  &    0.04 \\% &   0.018       \\
			       &  10\% samp.+\xspace\xspace2K   &   0.09  &  0.14 & 0.31 &   0.93   &  1.4   &  0.27   &  1.23  &  0.70   &   0.56   \\
			       &  10\% samp.+2.5K               &   0.09  &  0.14 & 0.31 &   0.17   &  0.30  &  0.09   &  1.46  &  0.19  &    0.04     \\ %
			       &  20\% samp.+\xspace\xspace2K   &   0.08  &  0.09 & 0.25 &   0.92   &  1.3   &  0.34   &  1.22  &  0.67  &    0.55    \\
			       &  20\% samp.+2.5K               &   0.08  &  0.09 & 0.25 &   0.12   &  0.36  &  0.04   &  1.52  & 0.19   &  0.05              \\ %
\hline \hline
\multirow{7}{*}{Epinions}  &  2K	           &    0    &   0   &    0  &   0.64   &  0.37   &  0.25  &  0.69  &  0.42   &    0.23        \\ %
	   	               &  2.25K            &    0    &   0   &    0  &   0.31   &  0.15   &  0.13  &  0.66  &  0.30   &    0.13   \\ %
                               &  2.5K             &    0    &   0   &    0  &   0.02   &  0.06   &  0.09  &  0.49  &  0.21   &    0.07    \\ %
			       &  10\% samp.+\xspace\xspace2K &   0.15  &  0.11 &  0.60 &  0.62    &  0.41   &  0.39  &  0.92  &  0.49   &    0.24      \\
			       &  10\% samp.+2.5K &  0.15 &  0.11 & 0.60  &  0.29    &  0.10   &  0.10  &  0.87  &  0.28   &    0.04         \\ %
			       &  20\% samp.+\xspace\xspace2K &    0.11 &  0.10 &  0.50 &  0.36    &  0.27   &  0.07  &  0.80  &  0.33   &    0.26    \\
			       &  20\% samp.+2.5K &  0.11 &  0.10 & 0.50  &  0.29   &   0.07   &  0.04  &  0.85  &  0.21   &    0.03       \\ %

\hline \hline 
\multirow{4}{*}{Enron}         &   2K                          &    0     &     0  &   0    &  0.73   & 1.01  &   0.16  &  1.21   &   0.94   &  0.24      \\
                               &  2.5K                         &    0     &     0  &   0    &  0.02   & 0.12  &  0.10   & 0.80    &   0.22   &  0.03 \\
			       &  10\% samp.+\xspace\xspace2K &     0.21  &  0.21  &   0.63 &  0.83   &  1.03 &  0.31   &  1.54   &   1.05   &  0.25       \\	
			       &  10\% samp.+2.5K             &    0.21   &  0.21  &   0.63 &  0.12   &  0.19  & 0.25   &  1.10   &   0.20   &  0.07        \\	
			       &  20\% samp.+\xspace\xspace2K &    0.23   &  0.22  &   0.62 &  0.78   & 1.06  &  0.30   &  1.42   &   0.92   &  0.16    \\	
			       &  20\% samp.+2.5K             &    0.23   &  0.22  &   0.62 &  0.11   & 0.12   &  0.12  &  1.50   &   0.19   &  0.08       \\	
\hline \hline
\multirow{5}{*}{\begin{minipage}{0.1in}CAIDA AS\end{minipage}}
                               &   2K                        &    0    &     0  &   0     &   0.44   &   0.22 &  0.26  &  0.23  &   0.52   &   0.06      \\
	   	               &  2.25K                       &    0    &     0  &   0     &   0.27   &   0.08 &  0.21  &  0.10  &   0.39   &   0.05          \\
                               &  2.5K                        &    0    &     0  &   0     &   0.02   &   0.03 &  0.08  &  0.03  &   0.38   &   0.04      \\
			       &  30\% samp.+\xspace\xspace2K &   0.12  &  0.28  &   0.49  &   0.58   &   0.32 &  0.24  &  0.32  &   0.73   &   0.05       \\	
			       &  30\% samp.+2.5K             &   0.12   &  0.28  &   0.49 &  0.31   &   0.15 &  0.18  &  0.26  &   0.39   &   0.05      \\	

\hline 
\end{tabular}}
  \caption{Results averaged over 5 runs. DD: Degree Distribution, Knn: Average Neighbor Degree Distribution, JDD: Joint degree Distribution, CC: Degree-Dependent Average Clustering, ESP: Edgewise shared partners, Sh.P.: Shortest Paths Distribution, Cliq.: Maximal Cliques Distribution, Cycl.: Cycle basis size distribution, Spect.: 20 largest eigenvalues.
  }
  \vspace{-20pt}
  \label{tab:results}
\end{table}

\begin{figure*}
\centering
\subfigure{
\includegraphics[scale=0.28, angle=0]{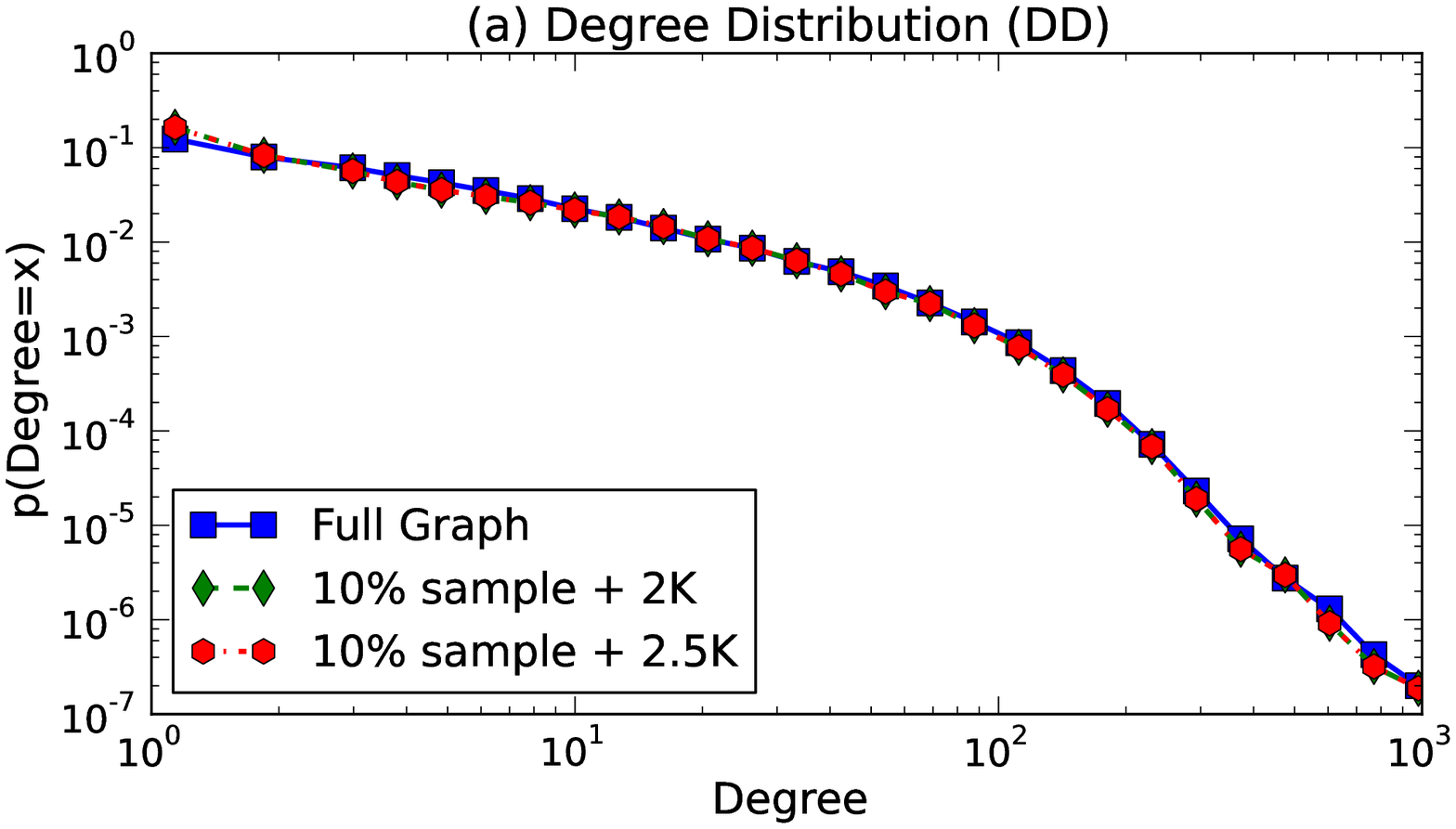}} 
\subfigure{
\includegraphics[scale=0.28, angle=0]{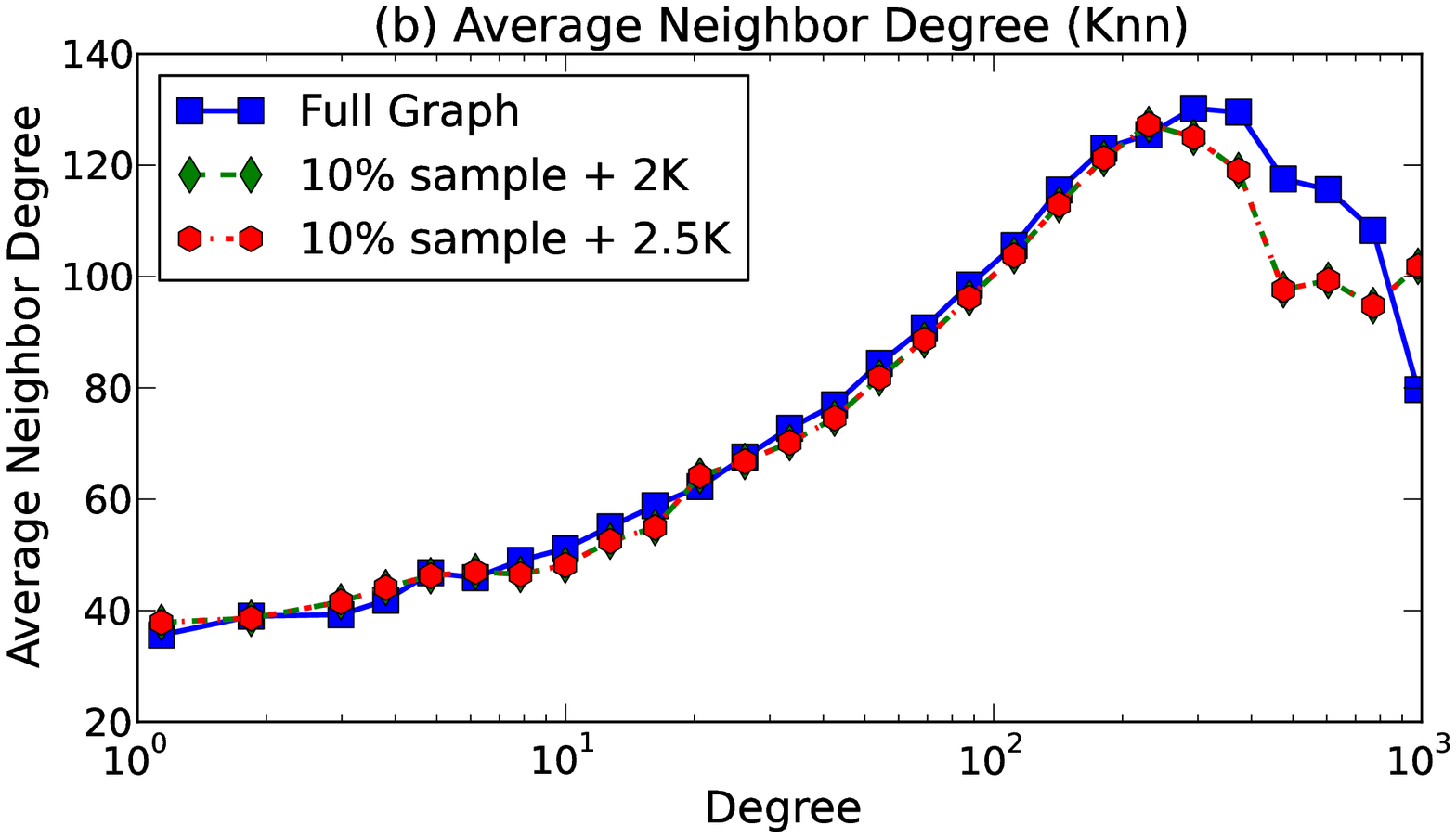}} 
\subfigure{
\includegraphics[scale=0.28, angle=0]{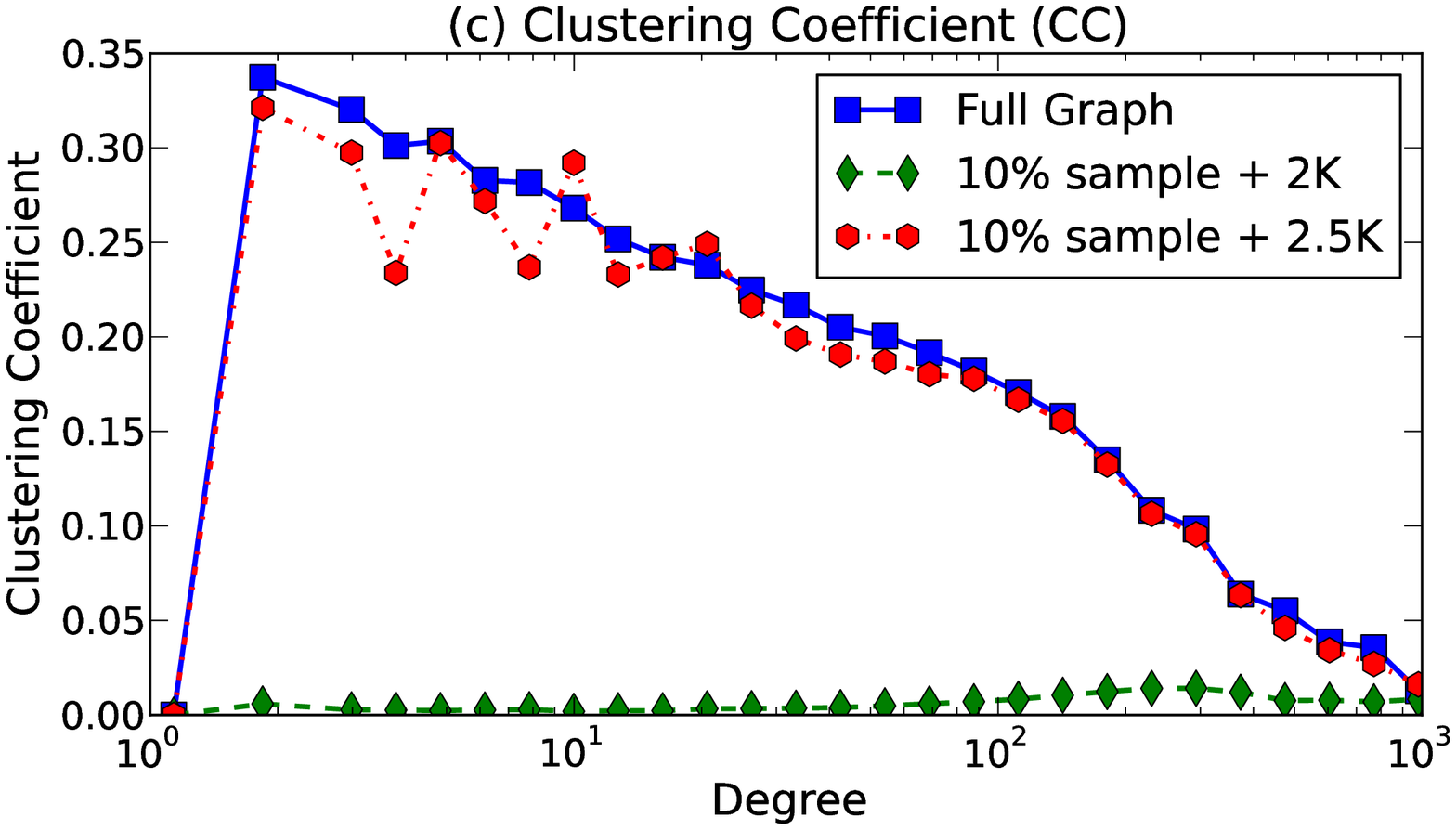}} 
\subfigure{
\includegraphics[scale=0.28, angle=0]{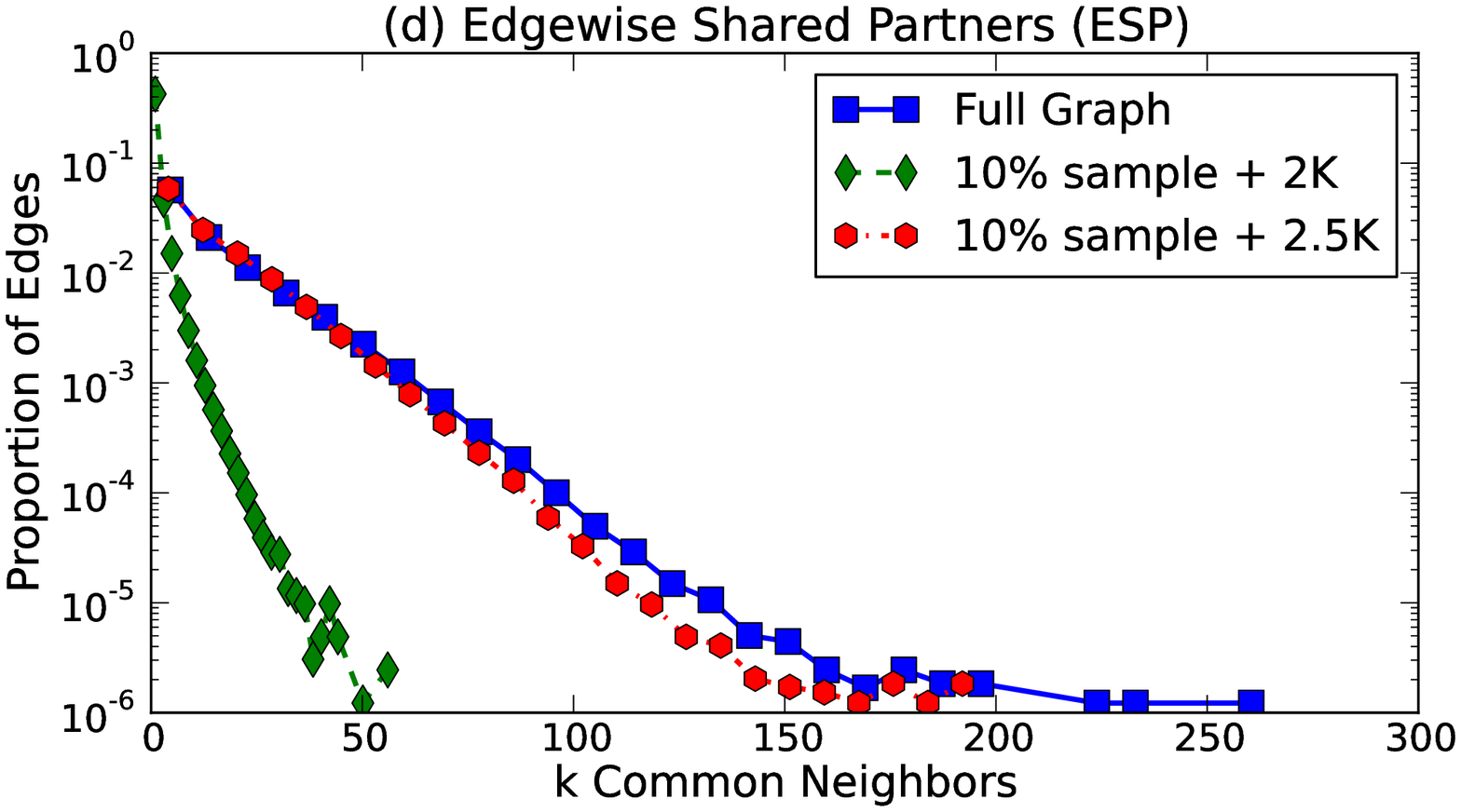}} 
\subfigure{
\includegraphics[scale=0.28, angle=0]{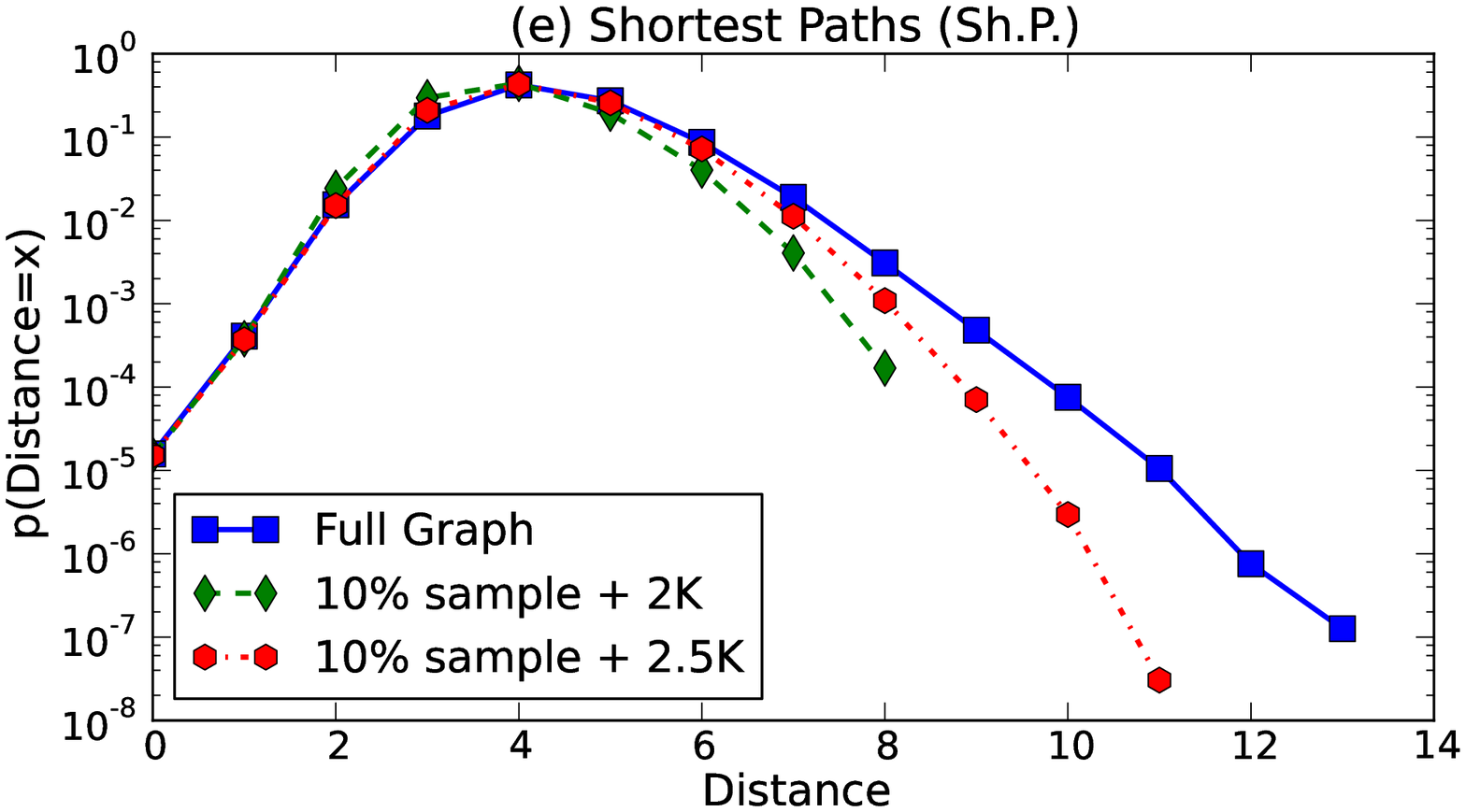}} 
\subfigure{
\includegraphics[scale=0.28, angle=0]{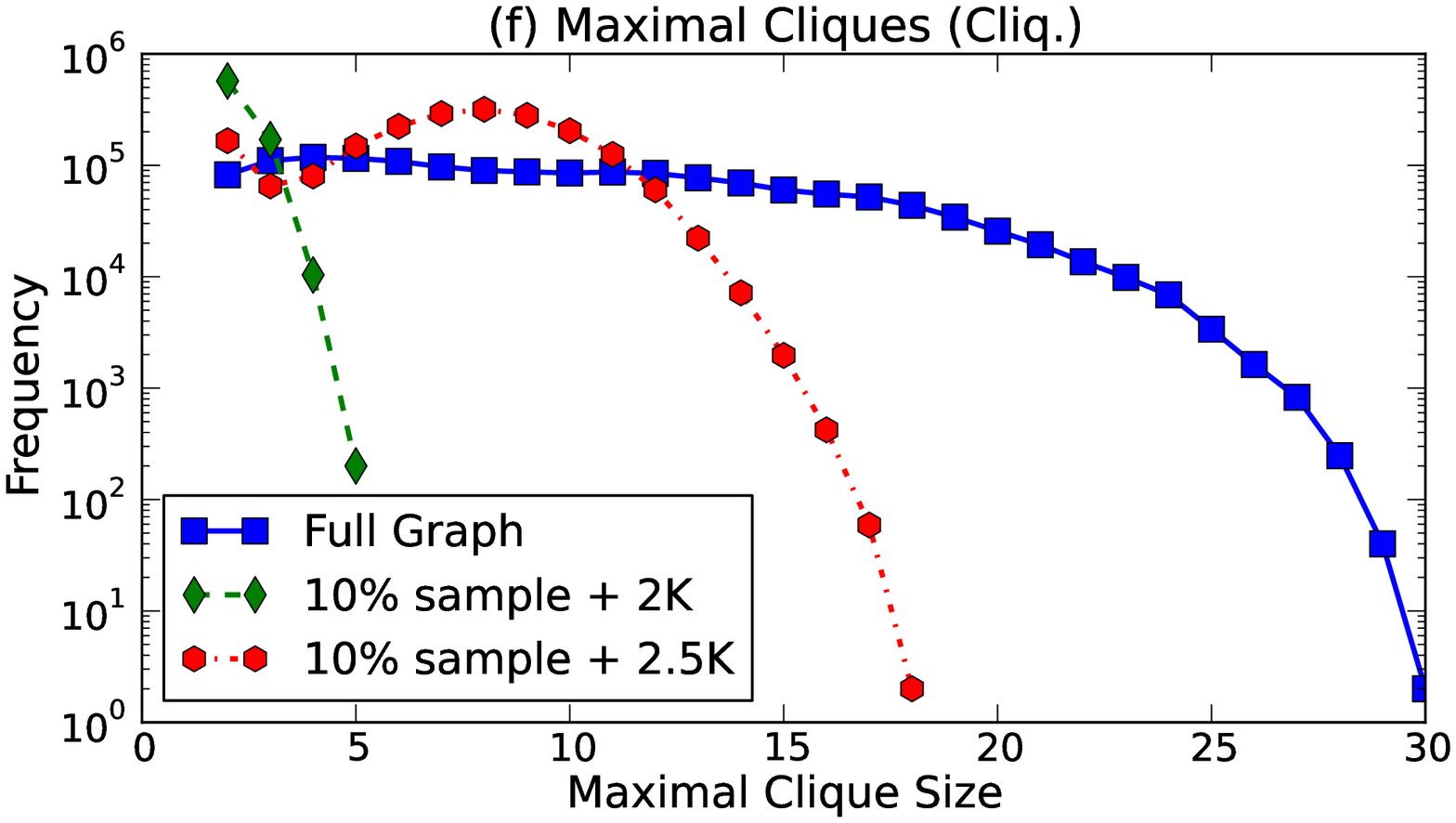}} 
\subfigure{
\includegraphics[scale=0.28, angle=0]{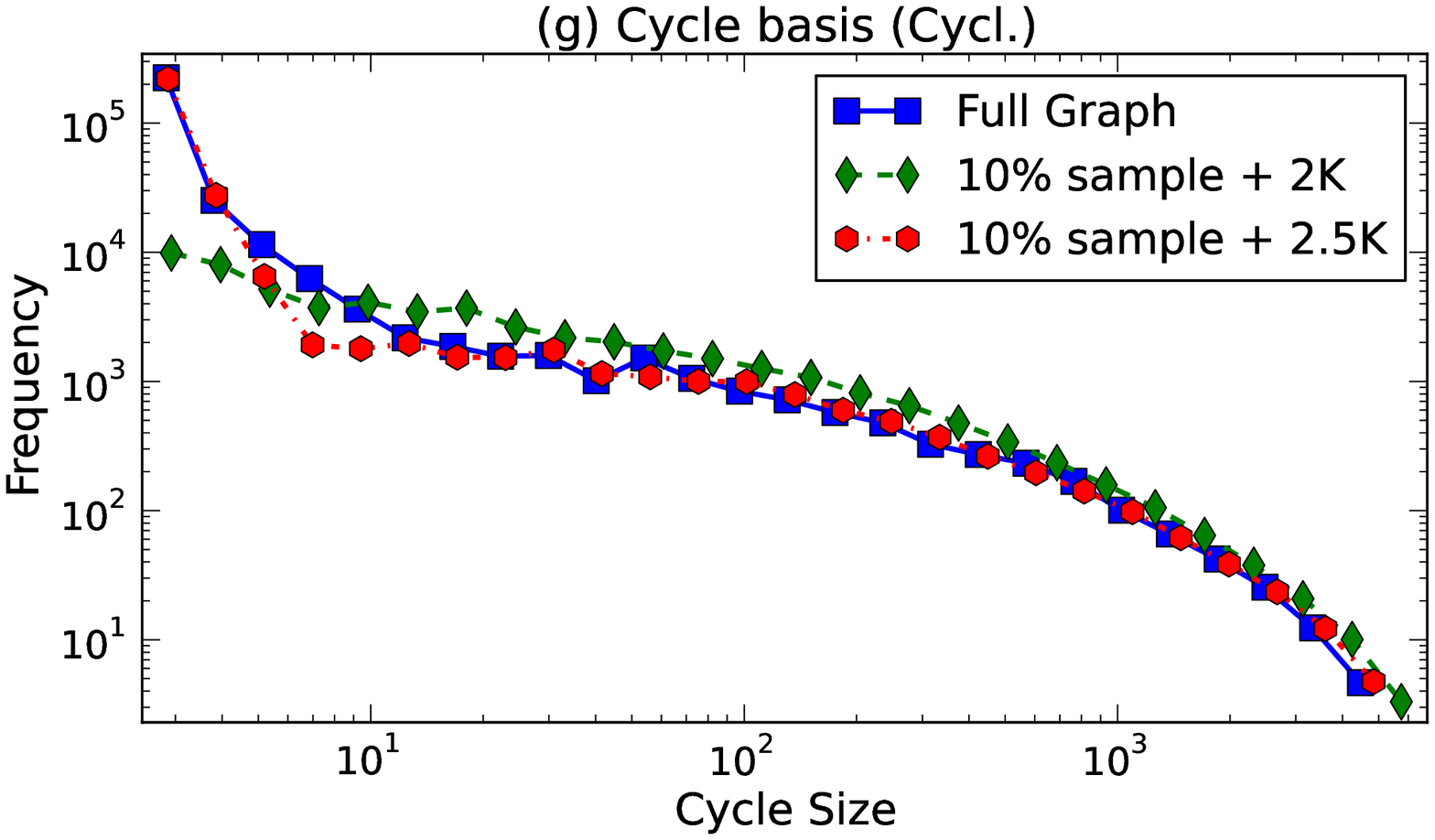}} 
\subfigure{
\includegraphics[scale=0.28, angle=0]{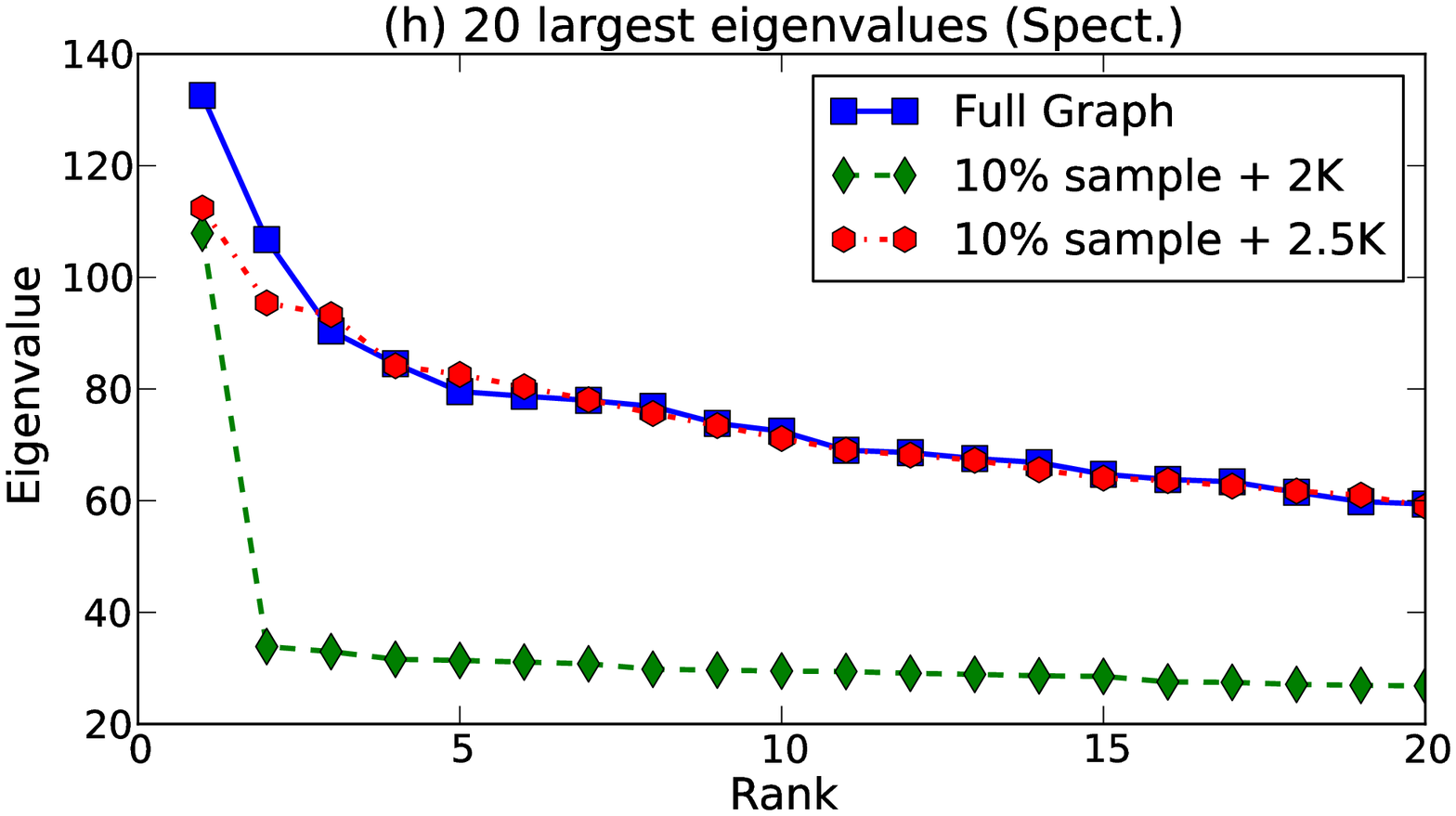}} 
\subfigure{
\includegraphics[scale=0.28, angle=0]{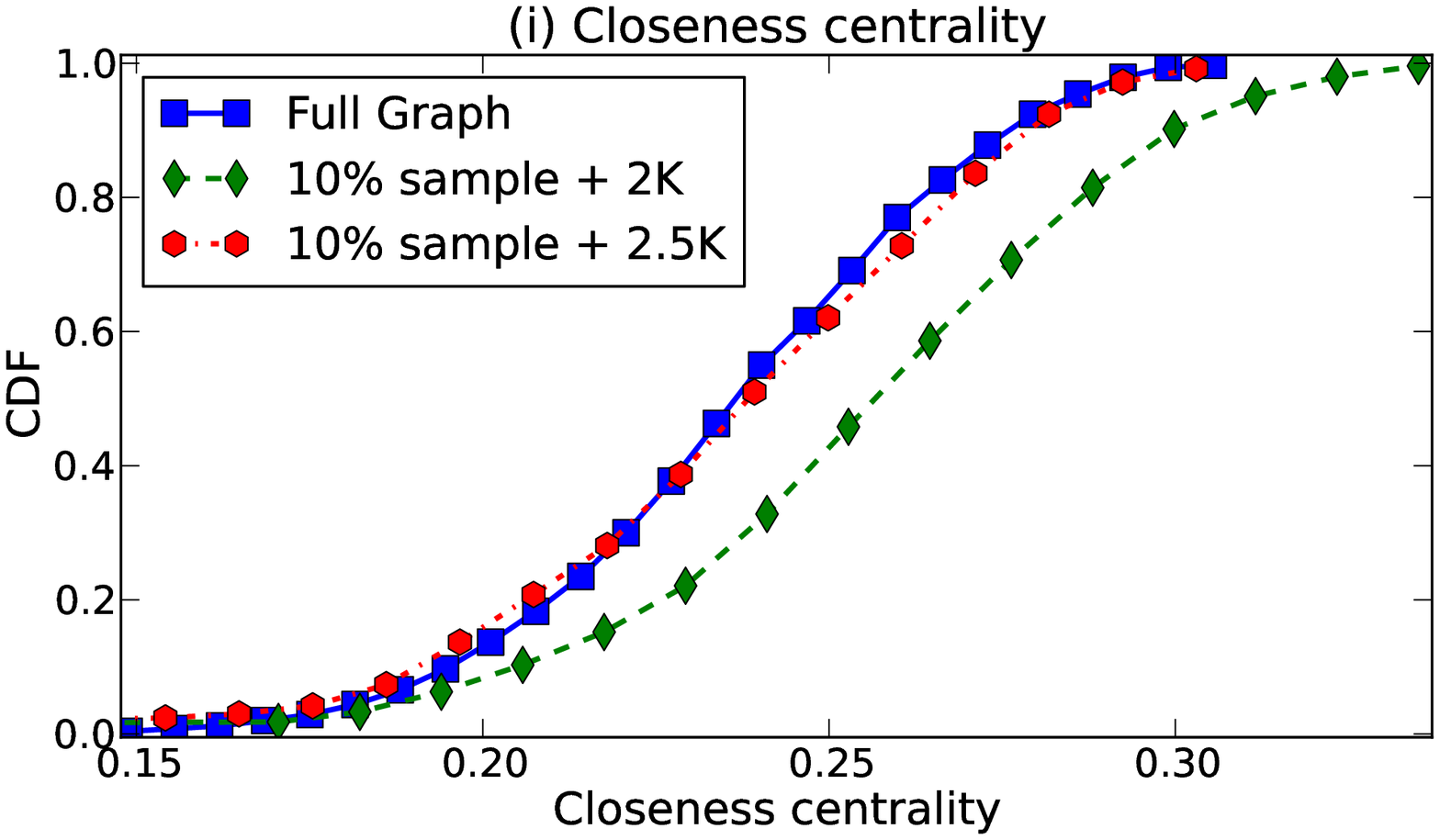}}
\caption{ {\em Facebook New Orleans} Distributions of nine graph properties for (i) the full graph (ii)  10\% RW sample + 2K construction (iii) 10\% RW sample + 2.5K construction. Results are binned in 30 intervals. }
\vspace{-15pt}
\label{fig:orleans_sample}
\end{figure*}

\subsubsection{Matching Graph Properties}

We  now compare how closely the generated graphs resembles the original ones, w.r.t. a  variety of graph properties. %
First, we present results  when the original graph is fully known, which provides a ground truth and allows to evaluate how close is the generated graph to the original.  Then, we also apply our methodology on unknown graphs, which we expect to be the main use in practice. 

\paragraph{Graph Properties used for Comparison}
We consider a range of graph properties, beyond just JDD and clustering which are met by construction. 
\\
(i) The degree distribution (DD) and degree-dependent average neighbor degree (Knn) are fully determined by $JDD(k,l)$. We include them mainly for the case in which the original graph is unknown and a sample is collected. \\
(ii) Edgewise shared partner distribution (ESP)
is the proportion of edges that have k common neighbors.  \\
(iii) Shortest path distribution (Sh.P.) is defined as the probability of a random pair of nodes to be at shortest path distance of h hops from each other. \\
(iv) Maximal clique distribution (Cliq.) is defined as the frequency of maximal cliques.  \\
(v) Cycles distribution (Cycl.) is the frequency of cycle length for a minimal cycle basis, in which a cycle cannot be re-constructed by the union of  cycles in the base. \\
(vi) Spectrum (Spect.): the eigenvalues of a graph are related to various graph properties such as expansion, and clusterability. \\
(vii) Closeness centrality of a node is defined as the inverse of the sum of distances of the node to all other nodes. It captures the speed that information spreads from this node to all other.

\paragraph{Results for a Fully Known Original Graph}

There are two natural questions regarding our approach:
\begin{itemize}
\item  {\em 2K vs. 2.5K:} how much does it help to target clustering in a graph, after having achieved $JDD$ exactly? 
\item {\em 2.25K vs 2.5K:} do we need to target the whole degree-dependent average clustering $\bar{c}(k)$ (2.5K) or can get most of the benefits by targeting the global average clustering coefficient $\bar{c}$ (2.25K)?
\end{itemize}

We answer these questions by performing experiments assuming that the graph is fully known. This allows us to examine the potential of our 2.5K generator without any estimation errors. For each known datasets we extract the exact $JDD(k,l)$, $\bar{c}$, $\bar{c}(k)$ and generate the $2K$, $2.25K$ and $2.5K$ graphs. 

We present the simulation results in \Tab{tab:results}. The results indicate that targeting $\bar{c}(k)$, on top of $JDD$, reduces the error on all considered graph properties, with the exception of the clique distribution. The graph properties that benefit the most are Spectrum, Shortest Path distribution, and Edgewise Shared Partners. Additionally, we deduce that $2.25K$ get as close as $2.5K$ to the Shortest Path distribution. However, on all other graph properties $2.5K$ is noticeably better when compared to $2.25K$ in terms of $NMAE$.

\paragraph{Results for Unknown Graph}

Finally, in this section we put all the pieces together. Our general work flow is the one shown in \Fig{fig:overview}): 
 (i)~sample the original graph $G$;  (ii)~estimate $JDD$ and $\bar{c}(k)$;  (iii)~post-process $JDD$; %%
 (iv)~apply our $2.5K$ generator to create a new graph $G'$; and  
 (v)~compare $G$ and $G'$ with respect to many metrics.

\Tab{tab:results} presents results for random walk samples of 10\%, and 20\% length  for the datasets New Orleans, Epinions, and Enron; samples of 20\%, and 30\% length for the smaller datasets UCSD, Harvard, 
and CAIDA AS. 
(We omit $2.25K$ since we have previously shown that $2.5K$ performs considerably better.) The results confirm that targeting $\bar{c}(k)$, in addition to achieving $JDD$, makes a big difference in terms of the $NMAE$ for all graph properties considered, despite the unavoidable measurement errors in the estimation of the model parameters. \Fig{fig:orleans_sample} shows plots of all considered graph properties for the New Orleans graph and compares between the full graph, a 10\% sample + 2K, and a 10\% sample + 2.5K. We observe that with just a 10\% sample we approximate extremely well the degree-dependent average clustering and the average neighbor degree. In addition, 2.5K gets much closer to all graph properties that were not targeted, with the exception of the maximal cliques.

\section{Conclusion}
\label{sec:conclusion}

Our work provides  a complete, and practical methodology for generating 2.5-K graphs that resemble a real (possibly unknown) graph. We present novel estimators, that measure our metrics of interest from node samples, and a novel 2.5K generator, that targets these metrics up to orders of magnitude faster than  prior approaches. We also make publicly available a Python implementation for all the building blocks  at \cite{2.5K_source_code}. We envision that an example application is the following: one can apply our methodology to construct graphs that resemble Facebook, without having access to the full social graph, by simply using crawling/sampling.

{
\small

\bibliographystyle{abbrv}
\vspace{-4pt}
\bibliography{OSN_Sampling_local,Topology_local,refs}
}

\end{document}